\documentclass{ws-mpla}
\usepackage[super]{cite}
\usepackage{graphicx}
\usepackage{hyperref}
\begin{document}

\def\xspace {}
\let\gev\undefined
\let\stat\undefined
\let\syst\undefined
\let\lum\undefined
\let\zp\undefined
\let\ga\undefined

\def\vdef #1{\expandafter\def\csname #1\endcsname}
\def\vuse #1{\csname #1\endcsname}
\def\vu   #1{\csname #1\endcsname}

\def\calc     {\ensuremath{{\cal C}}}
\def\tmm     {\ensuremath{\tau_{\mup\mun}}}
\def\cl       {\ensuremath{\,\text{CL}}}
\def\cls      {\ensuremath{\,\text{CL}_S}}
\def\mindz       {\ensuremath{\min|\Delta z|}}
\def\bdtvar      {\ensuremath{{\cal A}}}
\def\pdl         {proper decay length}
\def\npv         {\ensuremath{N_{\text{PV}}}}
\def\ndf         {\ensuremath{{\text{dof}}}}
\def\ndof        \ndf
\def\dof         \ndf
\def\pthat       {\ensuremath{\hat{p}_\perp}}
\def\Emax        {\ensuremath{E_{\text{max}}}}
\def\rmm         {\ensuremath{\Delta R(\mu\mu)}}
\def\ptmu        {\ensuremath{p_{\perp}^{\mu}}}
\def\ptmm        {\ensuremath{p_\perp^{\mu^+\mun}}}
\def\pmm         {\ensuremath{p^{\mu^+\mun}}}
\def\pttrk       {\ensuremath{p_\perp^{\mathrm{trk}}}}
\def\ptmuone     {\ensuremath{p_{\perp}^{\mu,1}}}
\def\ptmutwo     {\ensuremath{p_{\perp}^{\mu,2}}}
\def\ptb         {\ensuremath{p_{\perp}^{B}}}
\def\etamu       {\ensuremath{\eta_{\mu}}}
\def\etamuone    {\ensuremath{\eta_{\mu,1}}}
\def\etamutwo    {\ensuremath{\eta_{\mu,2}}}
\def\etab        {\ensuremath{\eta_{B}}}
\def\etamm       {\ensuremath{\eta^{\mu^+\mu^-}}}
\def\pvw         {\ensuremath{\langle w\rangle}}
\def\fl          {\ensuremath{\ell_{3D}}}
\def\fle         {\ensuremath{\sigma(\ell_{3D})}}
\def\flxy        {\ensuremath{\ell_{xy}}}
\def\fls         {\ensuremath{\ell_{3D}/\sigma(\ell_{3D})}}
\def\flsxy       {\ensuremath{\ell_{xy}/\sigma(\ell_{xy})}}
\def\pa          {\ensuremath{\alpha_{3\text{D}}}}
\def\chidof      {\ensuremath{\chi^2/\ndf}}
\def\closetrk    {\ensuremath{N_{\text{trk}}^{\text{close}}}}
\def\closetrkA   {\ensuremath{N_{\text{trk}}^{\text{close}, 1\sigma}}}
\def\closetrkB   {\ensuremath{N_{\text{trk}}^{\text{close}, 2\sigma}}}
\def\closetrkC   {\ensuremath{N_{\text{trk}}^{\text{close}, 3\sigma}}}
\def\iso         {\ensuremath{I}}
\def\isomu       {\ensuremath{I_{\mu}}}
\def\isomuone    {\ensuremath{I_{\mu_1}}}
\def\isomutwo    {\ensuremath{I_{\mu_2}}}
\def\xpdistmu    {\ensuremath{d_{\mu}^{xp}}}
\def\xpdistmuone {\ensuremath{d_{\mu_1}^{xp}}}
\def\xpdistmutwo {\ensuremath{d_{\mu_2}^{xp}}}
\def\docatrk     {\ensuremath{d^{\text{0}}_\text{ca}}}
\def\dca         {\ensuremath{d_\text{ca}}}
\def\dcasv       {\ensuremath{d_\text{ca}^\text{SV}}}
\def\maxdoca     {\ensuremath{d_\text{ca}^\text{max}}}
\def\ip          {\ensuremath{\delta_{3D}}}
\def\ips         {\ensuremath{\delta_{3D}/\sigma(\delta_{3D})}}
\def\lip         {\ensuremath{\ell_{z}}}
\def\lips        {\ensuremath{\ell_{z}/\sigma(\ell_{z})}}
\def\liptwo      {\ensuremath{\ell_{z}^{(2)}}}
\def\lipstwo     {\ensuremath{\ell_{z}^{(2)}/\sigma(\ell_{z}^{(2)})}}
\def\pvip        {\ensuremath{\delta_{3D}}}
\def\pvips       {\ensuremath{\delta_{3D}/\sigma(\delta_{3D})}}
\def\pvlip       {\ensuremath{\ell_{z}}}
\def\pvlips      {\ensuremath{\ell_{z}/\sigma(\ell_{z})}}
\def\pvliptwo    {\ensuremath{\ell_{z}^{(2)}}}
\def\pvlipstwo   {\ensuremath{\ell_{z}^{(2)}/\sigma(\ell_{z}^{(2)})}}
\def\pvdchi      {\ensuremath{\Delta(\chi^2)}}
\def\othervtx    {\ensuremath{\Delta{\mathcal P}}}

\def\mll   {\ensuremath{m_{\mu\mu}}}
\def\dof   {\ensuremath\text{dof}}

\def\ket#1        {\ensuremath|{#1}\rangle}
\def\bra#1        {\ensuremath\langle{#1}|}
\def\braket#1#2   {\ensuremath\langle{#1}|{#2}\rangle}
\def\tfi#1#2#3    {\ensuremath\langle{#1}|{#2}|{#3}\rangle}
\def\vtwo#1#2     {\ensuremath\left(\begin{array}{c}{#1}\\{#2}\end{array}\right)}
\def\vthree#1#2#3 {\ensuremath\left(\begin{array}{c}{#1}\\{#2}\\{#3}\end{array}\right)}
\def\me           {\ensuremath\mathcal{M}}
\def\ame          {\ensuremath|\mathcal{M}|^2}
\def\asme         {\ensuremath\overline{|\mathcal{M}|^2}}

\def\psib         {\ensuremath\overline{\psi}}

\newcommand{\mat}[2][cccccccccccccccccccccccccc]{\left(
   \begin{array}{#1}
    #2\\
   \end{array}
  \right)
}

\def\vvA        {\ensuremath\left(\begin{matrix} 0 \\ 0 \end{matrix}\right)}
\def\vvB        {\ensuremath\left(\begin{matrix} 1 \\ 0 \end{matrix}\right)}
\def\vvC        {\ensuremath\left(\begin{matrix} 0 \\ 1 \end{matrix}\right)}
\def\vvD        {\ensuremath\left(\begin{matrix} 1 \\ 1 \end{matrix}\right)}

\def\vvvA        {\ensuremath\left(\begin{matrix} 0 \\ 0 \\ 0\end{matrix}\right)}
\def\vvvB        {\ensuremath\left(\begin{matrix} 1 \\ 0 \\ 0\end{matrix}\right)}
\def\vvvC        {\ensuremath\left(\begin{matrix} 0 \\ 1 \\ 0\end{matrix}\right)}
\def\vvvD        {\ensuremath\left(\begin{matrix} 0 \\ 0 \\ 1\end{matrix}\right)}

\def\cpt   {\ensuremath{C\kern-0.2em P\kern-0.1em T}}
\def\cp    {\ensuremath{C\kern-0.2em P}}
\def\cpv   {\ensuremath{C\kern-0.2em P\kern-1.0em / }}
\def\CPV   {\cp-violation}
\def\CPTV  {\cpt-violation}
\def\bfsx  {$B$-physics}
\def\ETm   {\ensuremath{E_T\kern-1.2em/\kern0.6em}}
\def\ET    {\ensuremath{E_T}}
\def\kT    {\ensuremath{k_T}}
\def\ptm   {\ensuremath{p_\perp\kern-1.1em/\kern0.5em}}
\def\pvecm {\ensuremath{\vec{p} \kern-0.4em/\kern0.1em}}
\def\pvec  {\ensuremath{\vec{p}}}

\def\dsj   {\ensuremath{D_{sJ}}}
\def\vxb   {\ensuremath{|V_{xb}|}}
\def\vud   {\ensuremath{|V_{ud}|}}
\def\vus   {\ensuremath{|V_{us}|}}
\def\vub   {\ensuremath{|V_{ub}|}}
\def\vcd   {\ensuremath{|V_{cd}|}}
\def\vcs   {\ensuremath{|V_{cs}|}}
\def\vcb   {\ensuremath{|V_{cb}|}}
\def\vtd   {\ensuremath{|V_{td}|}}
\def\vts   {\ensuremath{|V_{ts}|}}
\def\vtb   {\ensuremath{|V_{tb}|}}

\def\deltam{\ensuremath{\delta m}}
\def\dm    {\ensuremath{\Delta m}}
\def\dt    {\ensuremath{\Delta t}}
\def\dg    {\ensuremath{\Delta \gamma}}
\def\dG    {\ensuremath{\Delta \Gamma}}
\def\dGs   {\ensuremath{\Delta \Gamma_s}}
\def\dmt   {\ensuremath{\Delta mt}}
\def\dmdt  {\ensuremath{\Delta m \Delta t}}
\def\dms   {\ensuremath{\Delta m_s}}
\def\dmst {\ensuremath{\Delta m_s t}}
\def\dmm   {\ensuremath{\Delta m^2}}
\def\TBY   {\ensuremath{\theta_{\Bz, D^*\ell}}}

\def\de    {\ensuremath{\Delta E}}
\def\mes   {\ensuremath{m_{ES}}}

\def\msd{\ensuremath{\overline{m}_D^2}}
\def\lbar{\ensuremath{\overline{\Lambda}}}
\def\lone{\ensuremath{\lambda_1}}
\def\ltwo{\ensuremath{\lambda_2}}

\def\MUP   {\ensuremath{\mu_\pi^2}}
\def\MUG   {\ensuremath{\mu_G^2}}
\def\RHOD  {\ensuremath{\rho_D^3}}
\def\RHOLS {\ensuremath{\rho_{LS}^3}}

\def\cd {\ensuremath{{\mathcal D}}}
\def\cbf {\ensuremath{{\mathcal B}}}
\def\cbfb{\ensuremath{{\overline{\mathcal B}}}}
\def\clu {\ensuremath{{\mathcal L}}}
\def\cor {\ensuremath{{\mathcal O}}}
\def\adg {\ensuremath{{\mathcal A}_{\Delta\Gamma}}}
\def\adgmm {\ensuremath{{\mathcal A}_{\Delta\Gamma}^{\mup\mun}}}

\def\mmiss{\ensuremath{{m_{miss}^2}}}
\def\rusl{\ensuremath{{R_{u}}}}
\def\mh{\ensuremath{{m_{had}}}}
\def\mmxx {\ensuremath{\langle m_X^2 \rangle~}}
\def\mmx {\ensuremath{\langle m_X \rangle~}}
\def\mxqq{\ensuremath{(m_X, Q^2)}}
\def\mX{\ensuremath{{m_X}}}
\def\mx{\ensuremath{{m_X}}}
\def\mxcut{\ensuremath{{m_X^{cut}}}}
\def\pstar{\ensuremath{{p^*}}}
\def\qtot{\ensuremath{{Q_{tot}}}}
\def\pt{\ensuremath{{p_\perp}}}
\def\mt{\ensuremath{{m_\perp}}}

\def\pslash{\ensuremath{{p\kern-0.45em /}}}
\def\pvecslash{\ensuremath{{\vec{p} \kern-0.45em /}}}

\def\meanmxx   {\ensuremath{\langle m_X^2 \rangle}}
\def\meanmx    {\ensuremath{\langle m_{X} \rangle}}
\def\mean#1    {\ensuremath{\langle #1 \rangle}}

\def\Bpilnu    {\ensuremath{\Bb\to \pi\ell\nub}}
\def\Bmpilnu   {\ensuremath{\Bm\to \pi^0\ell^-\nub}}
\def\Betalnu   {\ensuremath{\Bb\to \eta\ell\nub}}
\def\Brholnu   {\ensuremath{\Bb\to \rho\ell\nub}}
\def\Bmrholnu  {\ensuremath{\Bb\to \rho^0\ell^-\nub}}
\def\Bomegalnu {\ensuremath{\Bb\to \omega\ell\nub}}
\def\Brhoenu   {\ensuremath{\Bb\to \rho e\nub}}
\def\Bzrhoenu  {\ensuremath{\Bz\to \rho^- e^+\nub}}

\def\Bdlnu     {\ensuremath{\Bb\to D\ell\nub}}
\def\Bdstarlnu {\ensuremath{\Bb\to \Dstar \ell \nub}}
\def\Bzdstarlnu {\ensuremath{\Bzb\to \Dstarp \ell^- \nub}}
\def\Bzdstarenu {\ensuremath{\Bzb\to \Dstarp e^- \nub}}

\def\bll     {\ensuremath{\Bz\to \ell^+\ell^-}}
\def\bee     {\ensuremath{\Bz\to e^+e^-}}
\def\bmm     {\ensuremath{B\to \mu^+\mu^-}}
\def\bem     {\ensuremath{\Bz\to e^\pm\mu^\mp}}
\def\btt     {\ensuremath{\B\to \tau^+\tau^-}}
\def\bdtt     {\ensuremath{\Bz\to \tau^+\tau^-}}
\def\bdet     {\ensuremath{\Bz\to e^\pm\tau^\mp}}
\def\bdmt     {\ensuremath{\Bz\to \mu^\pm\tau^\mp}}
\def\bstt     {\ensuremath{\Bs\to \tau^+\tau^-}}
\def\bsmt     {\ensuremath{\B\to \mu^\pm\tau^\mp}}
\def\bmt     {\ensuremath{\B\to \mu^\pm\tau^\mp}}
\def\bmn     {\ensuremath{B^+\to \mu^+\nu_\mu}}
\def\btn     {\ensuremath{B^+\to \tau^+\nu_\tau}}
\def\bln     {\ensuremath{B^+\to \ell^+\nu_\ell}}
\def\bgen    {\ensuremath{B^-\to \gamma e\nub}}
\def\bgee    {\ensuremath{B^-\to \gamma e^+e^-}}
\def\bgg     {\ensuremath{B^-\to \gamma \gamma}}

\def\jpsitomu {\ensuremath{\jpsi\to \mu^+\mu^-}}
\def\bdmm     {\ensuremath{B^0\to  \mu^+\mu^-}}
\def\bhh      {\ensuremath{B\to  h^+h^-}}
\def\bhmunu   {\ensuremath{B\to  h^-\mup\nu}}
\def\bhmumu   {\ensuremath{B\to  h\mup\mun}}
\def\bdpipi   {\ensuremath{B^0\to  \pip\pim}}
\def\bdpik    {\ensuremath{B^0\to  \Kp\pim}}
\def\bdpimunu {\ensuremath{B^0\to  \pim\mup\nu}}
\def\bdpimumu  {\ensuremath{B^0\to \piz\mup\mun}}
\def\bdmumupz {\ensuremath{B^0\to  \mup\mun\piz}}
\def\bupimumu  {\ensuremath{B^-\to \pim\mup\mun}}
\def\lbppi    {\ensuremath{\Lambda_b\to p \pim }}
\def\lbpk     {\ensuremath{\Lambda_b\to p K^-}}
\def\lbpmunu  {\ensuremath{\Lambda_b\to p\mun\nub}}
\def\butrmunu {\ensuremath{B^+\to \mup\mun\mup \nu}}
\def\bctrmunu {\ensuremath{B_c\to \mup\mun\mup \nu}}
\def\bcpsimunu{\ensuremath{B_c\to \jpsi\mup \nu}}

\def\bdmm     {\ensuremath{B^0  \to \mu^+\mu^-}}
\def\bsmm     {\ensuremath{B^0_s\to \mu^+\mu^-}}
\def\bqmm     {\ensuremath{B^0_{q}\to \mu^+\mu^-}}
\def\bxmm     {\ensuremath{B^0_s(5.1\gev)\to \mu^+\mu^-}}
\def\bymm     {\ensuremath{B^0_s(5.7\gev)\to \mu^+\mu^-}}
\def\bskk     {\ensuremath{B^0_s\to K^+K^-}}
\def\bspipi   {\ensuremath{B^0_s\to \pip\pim}}
\def\bspik    {\ensuremath{B^0_s\to K^-\pip}}
\def\bskpi    {\ensuremath{B^0_s\to K^-\pip}}
\def\bdkpi    {\ensuremath{B^0\to K^+\pim}}
\def\bdkk     {\ensuremath{B^0\to K^+K^-}}
\def\bdpipi   {\ensuremath{B^0\to \pip\pim}}
\def\bdpik    {\ensuremath{B^0\to K^-\pip}}
\def\bskmunu  {\ensuremath{B^0_s\to K^-\mup\nu}}
\def\bsmumug  {\ensuremath{B^0_s\to \mup\mun\gamma}}
\def\bsmmg    {\ensuremath{B^0_s\to \mu^+\mu^-\gamma}}
\def\bdll     {\ensuremath{B^0\to \ell^+\ell^-}}
\def\bsll     {\ensuremath{B^0_s\to \ell^+\ell^-}}
\def\bstt     {\ensuremath{B^0_s\to \tau^+\tau^-}}
\def\bdmt     {\ensuremath{B^0  \to \mu^\pm\tau^\mp}}
\def\bsmt     {\ensuremath{B^0_s\to \mu^\pm\tau^\mp}}
\def\bsdmm    {\ensuremath{B^0_{s (d)}\to \mu^+\mu^-}}
\def\bszmm    {\ensuremath{B^0_{(s)}\to \mu^+\mu^-}}
\def\bsdll    {\ensuremath{B^0_{s (d)}\to \ell^+\ell^-}}
\def\bsdtt    {\ensuremath{B^0_{s (d)}\to \tau^+\tau^-}}
\def\tmmm     {\ensuremath{\tau\to \mu\mu\mu}}
\def\zjpsill  {\ensuremath{\Z\to \jpsi\ell^+\ell^-}}

\def\hbb      {\ensuremath{H\to \bbbar}}
\def\htt      {\ensuremath{H\to \taup\taum}}
\def\ttH      {\ensuremath{\ttbar H}}

\def\meg      {\ensuremath{\mu\to e\gamma}}
\def\meee     {\ensuremath{\mu\to eee}}
\def\pgg      {\ensuremath{\piz\to \g\g}}

\def\bsg     {\ensuremath{b\to s\gamma}}
\def\bulnu   {\ensuremath{b\to u\ell\nub}}
\def\bclnu   {\ensuremath{b\to c\ell\nub}}
\def\bcenu   {\ensuremath{b\to c e\nub}}
\def\bcmunu  {\ensuremath{b\to c \mu\nub}}
\def\buenu   {\ensuremath{b\to u e\nub}}

\def\bxlnu   {\ensuremath{b\to X\ell^-\nub}}
\def\Bxenu   {\ensuremath{\Bb\to Xe^-\nu}}

\newcommand {\Bxlnu}{\ensuremath{\Bb \rightarrow X \ell \bar{\nu}}}
\newcommand {\Bxclnu}{\ensuremath{\Bb \rightarrow X_c \ell \bar{\nu}}}
\newcommand {\Bxulnu}{\ensuremath{\Bb \rightarrow X_u \ell \bar{\nu}}}
\def\Bpxenu {\ensuremath{\Bp\to Xe^+\nu}}
\def\Bzxenu {\ensuremath{\Bz\to Xe^+\nu}}

\def\Bulnu   {\ensuremath{\Bb\to X_{u}\ell\nub}}
\def\Buenu   {\ensuremath{\Bb\to X_{u} e\nub}}
\def\Bclnu   {\ensuremath{\Bb\to X_{c}\ell\nub}}
\def\Bxulnu  {\ensuremath{\Bb\to X_{u}\ell\nub}}
\def\Bxuenu  {\ensuremath{\Bb\to X_{u} e\nub}}
\def\Bxclnu  {\ensuremath{\Bb\to X_{c}\ell\nub}}

\def\bfactory  {{{\sl B}-factory}}
\def\bfactories{{{\sl B}-factories}}
\def\bFactory  {{{\sl B}-Factory}}
\def\breco     {\ensuremath{B_{reco}}}
\def\btag      {\ensuremath{B_{tag}}}
\def\bdecay    {{$B$-decay}}
\def\bDecay    {{$B$-Decay}}
\def\bdecays   {{$B$-decays}}
\def\bDecays   {{$B$-Decays}}
\def\bhadron   {{$b$-hadron}}
\def\bhadrons  {{$B$-hadrons}}
\def\bmeson    {{$B$-meson}}
\def\bmesons   {{$B$-mesons}}
\def\bquark    {{$b$-quark}}
\def\bquarks   {{$b$-quarks}}
\def\bphysics  {{$b$-physics}}
\def\Bphysics  {{$B$-physics}}

\def\ie   {{\it i.e.}}
\def\cf   {{\it cf.}}
\def\eg   {{\it e.g.}}
\def\etal {{\it et~al.}}
\def\etc  {{\it etc.}}

\def\rtr    {{$\red\triangleright\black$}}
\def\barrow {{$\blue\to\black$}}
\def\bpoint {{$\blue\bullet\black$}}
\def\npoint {{\phantom{$ \blue\bullet\black$}}}

\newcommand\bfac   {$B$-Factories}

\newcommand\bu   {\ensuremath{b\to u}}
\newcommand\bc   {\ensuremath{b\to c}}

\newcommand\islbcd {inclusive semileptonic $B\to c\ell\nu$}
\newcommand\islbud {inclusive semileptonic $B\to u\ell\nu$}

\def\tg     {\ensuremath {\theta^{*}_T}}
\def\ctg     {\ensuremath {\cos{\tg}}}
\def\cth     {\ensuremath {\cos{\theta_{H}}}}
\def\cthe    {\ensuremath {\cos{\theta_{H\,\eta'}}}}
\def\cthr    {\ensuremath {\cos{\theta_{H\,\rho}}}}
\def\ctb     {\ensuremath {\cos{\theta^{*}_{B}}}}
\def\ebeam     {\ensuremath {E^{*}_{b}}}
\def\egcms     {\ensuremath {E^{*}_{\gamma}}}
\def\mkpi      {\ensuremath {M_{\Kp \pim}}}


\let\emi\en
\def\electron   {\ensuremath{e}\xspace}
\def\en         {\ensuremath{e^-}\xspace}   
\def\ep         {\ensuremath{e^+}\xspace}
\def\epm        {\ensuremath{e^\pm}\xspace}
\def\epem       {\ensuremath{e^+e^-}\xspace}
\def\ee         {\ensuremath{e^-e^-}\xspace}

\def\mmu        {\ensuremath{\mu}\xspace}
\def\mup        {\ensuremath{\mu^+}\xspace}
\def\mun        {\ensuremath{\mu^-}\xspace} 
\def\mumu       {\ensuremath{\mu^+\mu^-}\xspace}
\def\mtau       {\ensuremath{\tau}\xspace}

\def\taup       {\ensuremath{\tau^+}\xspace}
\def\taum       {\ensuremath{\tau^-}\xspace}
\def\tautau     {\ensuremath{\tau^+\tau^-}\xspace}

\def\ellm       {\ensuremath{\ell^-}\xspace}
\def\ellp       {\ensuremath{\ell^+}\xspace}
\def\ellell     {\ensuremath{\ell^+ \ell^-}\xspace}

\def\ellb        {\ensuremath{\bar{\ell}}\xspace}
\def\nub        {\ensuremath{\bar{\nu}}\xspace}
\def\nunub      {\ensuremath{\nu{\bar{\nu}}}\xspace}
\def\nunub      {\ensuremath{\nu{\bar{\nu}}}\xspace}
\def\nue        {\ensuremath{\nu_e}\xspace}
\def\nueb       {\ensuremath{\nub_e}\xspace}
\def\nuenueb    {\ensuremath{\nue\nueb}\xspace}
\def\num        {\ensuremath{\nu_\mu}\xspace}
\def\numb       {\ensuremath{\nub_\mu}\xspace}
\def\numnumb    {\ensuremath{\num\numb}\xspace}
\def\nut        {\ensuremath{\nu_\tau}\xspace}
\def\nutb       {\ensuremath{\nub_\tau}\xspace}
\def\nutnutb    {\ensuremath{\nut\nutb}\xspace}
\def\nul        {\ensuremath{\nu_\ell}\xspace}
\def\nulb       {\ensuremath{\nub_\ell}\xspace}
\def\nulnulb    {\ensuremath{\nul\nulb}\xspace}


\def\g     {\ensuremath{\gamma}\xspace}
\def\gaga  {\ensuremath{\gamma\gamma}\xspace}  
\def\ggstar{\ensuremath{\gamma\gamma^*}\xspace}

\def\ega    {\ensuremath{e\gamma}\xspace}
\def\game   {\ensuremath{\gamma e^-}\xspace}
\def\epemg  {\ensuremath{e^+e^-\gamma}\xspace}


\def\H      {\ensuremath{H^0}\xspace}
\def\Hp     {\ensuremath{H^+}\xspace}
\def\Hm     {\ensuremath{H^-}\xspace}
\def\Hpm    {\ensuremath{H^\pm}\xspace}
\def\W      {\ensuremath{W}\xspace}
\def\Wp     {\ensuremath{W^+}\xspace}
\def\Wm     {\ensuremath{W^-}\xspace}
\def\Wpm    {\ensuremath{W^\pm}\xspace}
\def\Z      {\ensuremath{Z^0}\xspace}


\def\q     {\ensuremath{q}\xspace}
\def\qbar  {\ensuremath{\overline q}\xspace}
\def\Qbar  {\ensuremath{\overline Q}\xspace}
\def\ffbar {\ensuremath{f\overline f}\xspace}
\def\qqbar {\ensuremath{q\overline q}\xspace}
\def\QQbar {\ensuremath{Q\overline Q}\xspace}
\def\u     {\ensuremath{u}\xspace}
\def\ubar  {\ensuremath{\overline u}\xspace}
\def\uubar {\ensuremath{u\overline u}\xspace}
\def\d     {\ensuremath{d}\xspace}
\def\dbar  {\ensuremath{\overline d}\xspace}
\def\ddbar {\ensuremath{d\overline d}\xspace}
\def\s     {\ensuremath{s}\xspace}
\def\sbar  {\ensuremath{\overline s}\xspace}
\def\ssbar {\ensuremath{s\overline s}\xspace}
\def\c     {\ensuremath{c}\xspace}
\def\cbar  {\ensuremath{\overline c}\xspace}
\def\ccbar {\ensuremath{c\overline c}\xspace}
\def\b     {\ensuremath{b}\xspace}
\def\bbar  {\ensuremath{\overline b}\xspace}
\def\bbbar {\ensuremath{b\overline b}\xspace}
\def\t     {\ensuremath{t}\xspace}
\def\tbar  {\ensuremath{\overline t}\xspace}
\def\tbar  {\ensuremath{\overline t}\xspace}
\def\ttbar {\ensuremath{t\overline t}\xspace}
\def\pbar  {\ensuremath{\overline p}\xspace}
\def\ppbar {\ensuremath{p\overline p}\xspace}


\def\piz   {\ensuremath{\pi^0}\xspace}
\def\pizs  {\ensuremath{\pi^0\mbox\,\rm{s}}\xspace}
\def\ppz   {\ensuremath{\pi^0\pi^0}\xspace}
\def\pip   {\ensuremath{\pi^+}\xspace}
\def\pim   {\ensuremath{\pi^-}\xspace}
\def\pipi  {\ensuremath{\pi^+\pi^-}\xspace}
\def\pipm  {\ensuremath{\pi^\pm}\xspace}
\def\pimp  {\ensuremath{\pi^\mp}\xspace}

\def\kaon  {\ensuremath{K}\xspace}
\def\Kbar  {\kern 0.2em\bar{\kern -0.2em K}{}\xspace}
\def\Kb    {\ensuremath{\Kbar}\xspace}
\def\Kz    {\ensuremath{K^0}\xspace}
\def\Kzb   {\ensuremath{\Kbar^0}\xspace}
\def\KzKzb {\ensuremath{\Kz \kern -0.16em \Kzb}\xspace}
\def\Kp    {\ensuremath{K^+}\xspace}
\def\Km    {\ensuremath{K^-}\xspace}
\def\Kpm   {\ensuremath{K^\pm}\xspace}
\def\Kmp   {\ensuremath{K^\mp}\xspace}
\def\KpKm  {\ensuremath{\Kp \kern -0.16em \Km}\xspace}
\def\KS    {\ensuremath{K^0_{\scriptscriptstyle S}}\xspace}
\def\KL    {\ensuremath{K^0_{\scriptscriptstyle L}}\xspace}
\def\Kstarz  {\ensuremath{K^{*0}}\xspace}
\def\Kstarzb {\ensuremath{\Kbar^{*0}}\xspace}
\def\Kstar   {\ensuremath{K^*}\xspace}
\def\Kstarb  {\ensuremath{\Kbar^*}\xspace}
\def\Kstarp  {\ensuremath{K^{*+}}\xspace}
\def\Kstarm  {\ensuremath{K^{*-}}\xspace}
\def\Kstarpm {\ensuremath{K^{*\pm}}\xspace}
\def\Kstarmp {\ensuremath{K^{*\mp}}\xspace}

\newcommand{\etapr}{\ensuremath{\eta^{\prime}}\xspace}


\def\Dbar    {\kern 0.2em\bar{\kern -0.2em D}{}\xspace}
\def\Db      {\ensuremath{\Dbar}\xspace}
\def\Dz      {\ensuremath{D^0}\xspace}
\def\Dzb     {\ensuremath{\Dbar^0}\xspace}
\def\DzDzb   {\ensuremath{\Dz {\kern -0.16em \Dzb}}\xspace}
\def\Dp      {\ensuremath{D^+}\xspace}
\def\Dm      {\ensuremath{D^-}\xspace}
\def\Dpm     {\ensuremath{D^\pm}\xspace}
\def\Dmp     {\ensuremath{D^\mp}\xspace}
\def\DpDm    {\ensuremath{\Dp {\kern -0.16em \Dm}}\xspace}
\def\Dstar   {\ensuremath{D^*}\xspace}
\def\Dstarb  {\ensuremath{\Dbar^*}\xspace}
\def\Dstarz  {\ensuremath{D^{*0}}\xspace}
\def\Dstarzb {\ensuremath{\Dbar^{*0}}\xspace}
\def\Dstarp  {\ensuremath{D^{*+}}\xspace}
\def\Dstarm  {\ensuremath{D^{*-}}\xspace}
\def\Dstarpm {\ensuremath{D^{*\pm}}\xspace}
\def\Dstarmp {\ensuremath{D^{*\mp}}\xspace}
\def\Ds      {\ensuremath{D^+_s}\xspace}
\def\Dsb     {\ensuremath{\Dbar^+_s}\xspace}
\def\Dss     {\ensuremath{D^{*+}_s}\xspace}

\newcommand{\dstr}{\ensuremath{\Dstar}\xspace}
\newcommand{\dstrstr}{\ensuremath{D^{**}}\xspace}
\newcommand{\dsp}{\ensuremath{\Dstarp}\xspace}
\newcommand{\dsm}{\ensuremath{\Dstarm}\xspace}
\newcommand{\dsz}{\ensuremath{\Dstarz}\xspace}

\def\B       {\ensuremath{B}\xspace}
\def\Bbar    {\kern 0.18em\bar{\kern -0.18em B}{}\xspace}
\def\Bb      {\ensuremath{\Bbar}\xspace}
\def\BB      {\ensuremath{B\Bbar}\xspace}
\def\Bz      {\ensuremath{B^0}\xspace}
\def\Bzb     {\ensuremath{\Bbar^0}\xspace}
\def\BzBzb   {\ensuremath{\Bz {\kern -0.16em \Bzb}}\xspace}
\def\BsBsb   {\ensuremath{\Bs {\kern -0.16em \Bsb}}\xspace}
\def\Bu      {\ensuremath{B^+}\xspace}
\def\Bub     {\ensuremath{B^-}\xspace}
\def\Bp      {\ensuremath{\Bu}\xspace}
\def\Bm      {\ensuremath{\Bub}\xspace}
\def\Bpm     {\ensuremath{B^\pm}\xspace}
\def\Bmp     {\ensuremath{B^\mp}\xspace}
\def\BpBm    {\ensuremath{\Bu {\kern -0.16em \Bub}}\xspace}
\def\Bd      {\ensuremath{B^0_d}\xspace}
\def\Bs      {\ensuremath{B^0_s}\xspace}
\def\Bc      {\ensuremath{B^+_c}\xspace}
\def\Lb      {\ensuremath{\Lambda_b}\xspace}
\def\Bsb     {\ensuremath{\Bzb_s}\xspace}
\def\Nz      {\ensuremath{M^0}\xspace}
\def\Nbar    {\kern 0.18em\bar{\kern -0.18em M}{}}
\def\Nzb     {\ensuremath{\Nbar^0}}
\def\NzNzb   {\ensuremath{\Nz {\kern -0.16em \Nzb}}}
\def\Nh      {\ensuremath{M_H}\xspace}
\def\Nl      {\ensuremath{M_L}\xspace}
\def\Nphys   {\ensuremath{\Nz_{phys}(t)}\xspace}
\def\Nbphys  {\ensuremath{\Nzb_{phys}(t)}\xspace}
\def\gh      {\ensuremath{\gamma_H}\xspace}
\def\gl      {\ensuremath{\gamma_L}\xspace}

\def\Bzd     {\ensuremath{B_d^0}\xspace}
\def\Bzs     {\ensuremath{B_s^0}\xspace}
\def\Bsd     {\ensuremath{B_{s(d)}^0}\xspace}
\def\BsH     {\ensuremath{B_{s,H}^0}\xspace}
\def\BsL     {\ensuremath{B_{s,L}^0}\xspace}
\def\BsLH    {\ensuremath{B_{s,L(H)}^0}\xspace}


\def\jpsi     {\ensuremath{{J\mskip -3mu/\mskip -2mu\psi\mskip 2mu}}\xspace}
\def\psitwos  {\ensuremath{\psi{(2S)}}\xspace}
\def\psiprpr  {\ensuremath{\psi(3770)}\xspace}
\def\etac     {\ensuremath{\eta_c}\xspace}
\def\chiczero {\ensuremath{\chi_{c0}}\xspace}
\def\chicone  {\ensuremath{\chi_{c1}}\xspace}
\def\chictwo  {\ensuremath{\chi_{c2}}\xspace}
\mathchardef\Upsilon="7107
\def\Y#1S{\ensuremath{\Upsilon{(#1S)}}\xspace}
\def\OneS  {\Y1S}
\def\TwoS  {\Y2S}
\def\ThreeS{\Y3S}
\def\FourS {\Y4S}
\def\FiveS {\Y5S}

\def\chic#1{\ensuremath{\chi_{c#1}}\xspace} 


\def\proton      {\ensuremath{p}\xspace}
\def\antiproton  {\ensuremath{\overline p}\xspace}
\def\neutron     {\ensuremath{n}\xspace}
\def\antineutron {\ensuremath{\overline n}\xspace}

\mathchardef\Deltares="7101
\mathchardef\Xi="7104
\mathchardef\Lambda="7103
\mathchardef\Sigma="7106
\mathchardef\Omega="710A

\def\Deltabar{\kern 0.25em\overline{\kern -0.25em \Deltares}{}\xspace}
\def\Lbar{\kern 0.2em\overline{\kern -0.2em\Lambda\kern 0.05em}\kern-0.05em{}\xspace}
\def\Sigbar{\kern 0.2em\overline{\kern -0.2em \Sigma}{}\xspace}
\def\Xibar{\kern 0.2em\overline{\kern -0.2em \Xi}{}\xspace}
\def\Obar{\kern 0.2em\overline{\kern -0.2em \Omega}{}\xspace}
\def\Xb{\kern 0.2em\overline{\kern -0.2em X}{}\xspace}

\def\X {\ensuremath{X}\xspace}


\def\BR         {{\ensuremath{\mathcal B}\xspace}}
\def\BRtauptoe  {\ensuremath{\BR(\taup \to \ep)}\xspace}
\def\BRtaumtoe  {\ensuremath{\BR(\taum \to \en)}\xspace}
\def\BRtauptomu {\ensuremath{\BR(\taup \to \mup)}\xspace}
\def\BRtaumtomu {\ensuremath{\BR(\taum \to \mun)}\xspace}


\newcommand{\etaprepp}{\ensuremath{\etapr \to \eta \pipi}\xspace}
\newcommand{\etaprrg} {\ensuremath{\etapr \to \rho^0 \g}\xspace}


\def\bdpsikstar {\ensuremath{\Bz \to \jpsi \Kstarz}\xspace}
\def\bspsiphi   {\ensuremath{\Bs \to \jpsi \phi}\xspace}
\def\bspsif     {\ensuremath{\Bs \to \jpsi f_0}\xspace}
\def\bsphiphi   {\ensuremath{\Bs \to \jpsi \phi}\xspace}
\def\lblmumu    {\ensuremath{\Lambda_b \to \mup\mun \Lambda}\xspace}
\def\bsphimm    {\ensuremath{\Bs \to \phi\mup\mun}\xspace}
\def\bdkmm      {\ensuremath{\Bz \to K\mup\mun}\xspace}
\def\bpsiks     {\ensuremath{\Bz \to \jpsi \KS}\xspace}
\def\bpsikst    {\ensuremath{\Bz \to \jpsi \Kstar}\xspace}
\def\bpsikl     {\ensuremath{\Bz \to \jpsi \KL}\xspace}
\def\bpsikzeropi{\ensuremath{\Bz \to \jpsi \Kstarz (\to \KL \piz)}\xspace}
\def\bpsikpluspi{\ensuremath{\Bu \to \jpsi \Kstarp (\to \KL \pip)}\xspace}
\def\bpsikpi    {\ensuremath{\Bz \to \jpsi (\to \mumu) \Kpm \pimp}\xspace}
\def\bupsik     {\ensuremath{\Bp \to \jpsi \Kp}\xspace}
\def\bupsikp    {\ensuremath{\Bu \to\jpsi\Kp}\xspace}
\def\bupsipi    {\ensuremath{\Bp \to \jpsi \pip}\xspace}
\def\bupsimmk   {\ensuremath{\Bp \to \jpsi (\to \mumu) \Kp}\xspace}
\def\bpsiX      {\ensuremath{\B \to \jpsi \X}\xspace}

\def\Bzbtomu    {\ensuremath{\Bzb \to \mu \X}\xspace}
\def\Bzbtox     {\ensuremath{\Bzb \to \X}\xspace}
\def\Bztopipi   {\ensuremath{\Bz \to \pipi}\xspace}
\def\Bztokpi    {\ensuremath{\Bz \to \Kp \pim}\xspace}
\def\Bztorhopi  {\ensuremath{\Bz \to \rho^+ \pim}\xspace}
\def\Bztorhorho {\ensuremath{\Bz \to \rho \rho}\xspace}
\def\Bztokrho   {\ensuremath{\Bz \to K \rho}\xspace}
\def\Bztokstpi  {\ensuremath{\Bz \to \Kstar \pi}\xspace}
\def\Bztoapi    {\ensuremath{\Bz \to a_1 \pi}\xspace}
\def\Bztodd     {\ensuremath{\Bz \to \DpDm}\xspace}
\def\Bztodstd   {\ensuremath{\Bz \to \Dstarp \Dm}\xspace}
\def\Bztodstdst {\ensuremath{\Bz \to \Dstarp \Dstarm}\xspace}
\def\Bztopimunu {\ensuremath{\Bs \to \pim \mup\nu}\xspace}

\def\Bstokk     {\ensuremath{\Bs \to K^+K^-}\xspace}
\def\Bstokpi    {\ensuremath{\Bs \to K^- \pip}\xspace}
\def\Bstokmunu  {\ensuremath{\Bs \to K^- \mup\nu}\xspace}

\def\BtoDK      {\ensuremath{B \to DK}\xspace}
\def\Btodstlnu  {\ensuremath{B \to \Dstar \ell \nu}\xspace}
\def\Btodstdlnu {\ensuremath{B \to \Dstar(D) \ell \nu}\xspace}
\def\Btorholnu  {\ensuremath{B \to \rho \ell \nu}\xspace}
\def\Btopilnu   {\ensuremath{B \to \pi \ell \nu}\xspace}

\def\Btoetah    {\ensuremath{B \to \eta h}\xspace}
\def\Btoetaph   {\ensuremath{B \to \etapr h}\xspace}

\newcommand{\Betaprks}{\ensuremath{\Bz \to \etapr \KS}\xspace}
\newcommand{\Betaprkz}{\ensuremath{\Bz \to \etapr \Kz}\xspace}

\def\btosgam    {\ensuremath{b \to s \g}\xspace}
\def\btodgam    {\ensuremath{b \to d \g}\xspace}
\def\btosll     {\ensuremath{b \to s \ellell}\xspace}
\def\btosmm     {\ensuremath{b \to s \mup\mun}\xspace}
\def\btosnunu   {\ensuremath{b \to s \nunub}\xspace}
\def\btosgaga   {\ensuremath{b \to s \gaga}\xspace}
\def\btosglue   {\ensuremath{b \to s g}\xspace}

\def\bmuX       {\ensuremath{b\to \mu X}\xspace}
\def\cmuX       {\ensuremath{c\to \mu X}\xspace}
\def\hmuX       {\ensuremath{h\to \mu X}\xspace}
\def\bbmumuX    {\ensuremath{\bbbar\to \mu^+\mu^-+X}\xspace}
\def\ccmumuX    {\ensuremath{\ccbar\to \mu^+\mu^-+X}\xspace}
\def\bpsimmX    {\ensuremath{b\to \jpsi(\to\mup\mun) X}\xspace}


\def\upsbb   {\ensuremath{\FourS \to \BB}\xspace}
\def\upsbzbz {\ensuremath{\FourS \to \BzBzb}\xspace}
\def\upsbpbm {\ensuremath{\FourS \to \BpBm}\xspace}
\def\upspsikl{\ensuremath{\FourS \to (\bpsikl) (\Bzbtox)}\xspace}


\def\tauptoe    {\ensuremath{\taup \to \ep \nunub}\xspace}
\def\taumtoe    {\ensuremath{\taum \to \en \nunub}\xspace}
\def\tauptomu   {\ensuremath{\taup \to \mup \nunub}\xspace}
\def\taumtomu   {\ensuremath{\taum \to \mun \nunub}\xspace}
\def\tauptopi   {\ensuremath{\taup \to \pip \nub}\xspace}
\def\taumtopi   {\ensuremath{\taum \to \pim \nu}\xspace}


\def\ggtopi     {\ensuremath{\gaga \to \pipi}\xspace}
\def\ggtopiz    {\ensuremath{\gaga \to \ppz}\xspace}
\def\ggstox     {\ensuremath{\ggstar \to \X(1420) \to \kaon \kaon \pi}\xspace}
\def\ggstoeta   {\ensuremath{\ggstar \to \eta(550) \to \pipi \piz}\xspace}


\def\ptot       {\mbox{$p$}\xspace}
\def\pxy        {\mbox{$p_T$}\xspace}
\def\pt         {\mbox{$p_\perp$}\xspace}
\def\ptrel      {\mbox{$p_\perp^{\text{rel}}$}\xspace}
\def\mes        {\mbox{$m_{\rm ES}$}\xspace}
\def\mec        {\mbox{$m_{\rm EC}$}\xspace}
\def\DeltaE     {\mbox{$\Delta E$}\xspace}

\def\pbcm {\ensuremath{p^*_{\Bz}}\xspace}


\def\mphi       {\mbox{$\phi$}\xspace}
\def\mtheta     {\mbox{$\theta$}\xspace}
\def\ctheta     {\mbox{$\cos\theta$}\xspace}

\newcommand{\ke}{\ensuremath{\mbox{\,ke}}\xspace}

\newcommand{\eev}{\ensuremath{\mbox{\,Ee\kern -0.1em V}}\xspace}
\newcommand{\pev}{\ensuremath{\mbox{\,Pe\kern -0.1em V}}\xspace}
\newcommand{\tev}{\ensuremath{\mbox{\,Te\kern -0.1em V}}\xspace}
\newcommand{\gev}{\ensuremath{\mbox{\,Ge\kern -0.1em V}}\xspace}
\newcommand{\mev}{\ensuremath{\mbox{\,Me\kern -0.1em V}}\xspace}
\newcommand{\kev}{\ensuremath{\mbox{\,ke\kern -0.1em V}}\xspace}
\newcommand{\ev} {\ensuremath{\mbox{\,e\kern -0.1em V}}\xspace}
\def\microEv         {\ensuremath{\,\mu\mbox{eV}}\xspace}  
\def\milliEv     {\ensuremath{\,\mbox{meV}}\xspace}  
\newcommand{\gevc}{\ensuremath{{\mbox{\,Ge\kern -0.1em V\!/}c}}\xspace}
\newcommand{\mevc}{\ensuremath{{\mbox{\,Me\kern -0.1em V\!/}c}}\xspace}
\newcommand{\gevcc}{\ensuremath{{\mbox{\,Ge\kern -0.1em V\!/}c^2}}\xspace}
\newcommand{\mevcc}{\ensuremath{{\mbox{\,Me\kern -0.1em V\!/}c^2}}\xspace}


\def\N   {\ensuremath{\mbox{\,N}\xspace}}

\def\syin {\ensuremath{^{\prime\prime}}\xspace}
\def\inch {\ensuremath{\rm \,in}\xspace} 
\def\ft   {\ensuremath{\rm \,ft}\xspace}
\def\km   {\ensuremath{\mbox{\,km}}\xspace}
\def\m    {\ensuremath{\mbox{\,m}}\xspace}
\def\cm   {\ensuremath{\mbox{\,cm}}\xspace}
\def\sr   {\ensuremath{\mbox{\,sr}}\xspace}
\def\cma  {\ensuremath{\mbox{\,cm}^2}\xspace}
\def\mm   {\ensuremath{\mbox{\,mm}\xspace}}
\def\mma  {\ensuremath{\mbox{\,mm}^2}\xspace}
\def\mum  {\ensuremath{\,\mu\mbox{m}\xspace}}
\def\muma {\ensuremath{\,\mu\mbox{m^2}}\xspace}
\def\fm   {\ensuremath{\mbox{\,fm}}\xspace}
\def\nm   {\ensuremath{\mbox{\,nm}}\xspace}   
\def\nb   {\ensuremath{\mbox{\,nb}}\xspace}
\def\barn      {\ensuremath{\mbox{\,b}}\xspace}
\def\mbarn     {\ensuremath{\mbox{\,mb}}\xspace}
\def\mb        {\ensuremath{\mbox{\,mb}}\xspace}
\def\pb        {\ensuremath{\mbox{\,pb}}\xspace}
\def\invmb     {\ensuremath{\mbox{\,mb}^{-1}}\xspace}
\def\invnb     {\ensuremath{\mbox{\,nb}^{-1}}\xspace}
\def\invpb     {\ensuremath{\mbox{\,pb}^{-1}}\xspace}
\def\ub        {\ensuremath{\,\mu\mbox{b}}\xspace}
\def\invub     {\ensuremath{\mbox{\,\ub}^{-1}}\xspace}
\def\fb        {\ensuremath{\mbox{\,fb}}\xspace}
\def\invfb     {\ensuremath{\mbox{\,fb}^{-1}}\xspace}
\def\ab        {\ensuremath{\mbox{\,ab}}\xspace}
\def\invab     {\ensuremath{\mbox{\,ab}^{-1}}\xspace}
\def\cms       {\ensuremath{\mbox{\,cm}^{-2}\mbox{s}^{-1}}\xspace}
\def\sqrts     {\ensuremath{\sqrt{s}}}

\def\mpc     {\ensuremath{\mbox{\,Mpc}}\xspace}

\def\kW   {\ensuremath{\mbox{\,kW}}\xspace}
\def\MW   {\ensuremath{\mbox{\,MW}}\xspace}
\def\mW   {\ensuremath{\mbox{\,mW}}\xspace}
\def\GW   {\ensuremath{\mbox{\,GW}}\xspace}

\def\mA   {\ensuremath{\mbox{\,mA}}\xspace}


\def\Hz  {\ensuremath{\mbox{\, Hz}}\xspace}
\def\kHz {\ensuremath{\mbox{\, kHz}}\xspace}
\def\MHz {\ensuremath{\mbox{\, MHz}}\xspace}

\def\us   {\ensuremath{\,\mu\mbox{s}}\xspace}
\def\ns   {\ensuremath{\mbox{\,ns}}\xspace}
\def\ms   {\ensuremath{\mbox{\,ms}}\xspace}
\def\ps   {\ensuremath{\mbox{\,ps}}\xspace}
\def\fs   {\ensuremath{\mbox{\,fs}}\xspace}
\def\gm   {\ensuremath{\mbox{\,g}}\xspace}

\def\Gy{\ensuremath{\mbox{\,Gy}}\xspace}
\def\sec{\ensuremath{\mbox{\,s}}\xspace}       
\def\msec{\ensuremath{\mbox{\,ms}}\xspace}       
\def\usec{\ensuremath{\,\mu \mbox{s}}\xspace}       
\def\h          {\ensuremath{\mbox{\,h}\xspace}}

\def\kg         {\ensuremath{\mbox{\,kg}}\xspace}  
\def\gram       {\ensuremath{\mbox{\,g}}\xspace}  

\def\uTesla     {\ensuremath{\,\mu\mbox{T}}\xspace}  
\def\mTesla     {\ensuremath{\mbox{\,mT}}\xspace}  
\def\nTesla     {\ensuremath{\mbox{\,nT}}\xspace}  
\def\Tesla      {\ensuremath{\mbox{\,T}}\xspace}  
\def\Gauss      {\ensuremath{\mbox{\,G}}\xspace}  

\def\mA     {\ensuremath{\mbox{\,mA}}\xspace}  
\def\Ampere     {\ensuremath{\mbox{\,A}}\xspace}  
\def\Amp     {\ensuremath{\mbox{\,A}}\xspace}  
\def\Watt     {\ensuremath{\mbox{\,W}}\xspace}  


\def\Xrad {\ensuremath{X_0}\xspace}
\def\NIL{\ensuremath{\lambda_{int}}\xspace}
\let\dgr\degrees

\def\mbar        {\ensuremath{\mbox{\,mbar}}}   

\def\C      {\ensuremath{\mbox{\, C}}\xspace}
\def\MVolts {\ensuremath{\mbox{\, MV}}\xspace}
\def\kVolts {\ensuremath{\mbox{\, kV}}\xspace}
\def\Volts {\ensuremath{\mbox{\, V}}\xspace}
\def\Volt  {\ensuremath{\mbox{\, V}}\xspace}
\def\atm   {\ensuremath{\mbox{\,atm}}\xspace}
\def\Ke    {\ensuremath{\mbox{\, K}}\xspace}
\def\mKe   {\ensuremath{\mbox{\, mK}}\xspace}
\def\mic  {\ensuremath{\,\mu{\rm C}}\xspace}
\def\krad {\ensuremath{\rm \,krad}\xspace}
\def\cmc  {\ensuremath{{\rm \,cm}^3}\xspace}
\def\yr   {\ensuremath{\rm \,yr}\xspace}
\def\hr   {\ensuremath{\rm \,hr}\xspace}
\def\degc {\ensuremath{^\circ}{C}\xspace}
\def\degk {\ensuremath{\,\mbox{K}}\xspace}
\def\degrees {\ensuremath{^{\circ}}\xspace}
\def\mrad {\ensuremath{\,\mbox{mrad}}\xspace}               
\def\urad {\ensuremath{\,\mu\mbox{rad}}\xspace}               
\def\rad{\ensuremath{\mbox{\,rad}}\xspace}
\def\mradhyph{\ensuremath{\rm -mr}\xspace}
\def\sx    {\ensuremath{\sigma_x}\xspace}
\def\sy    {\ensuremath{\sigma_y}\xspace}
\def\sz    {\ensuremath{\sigma_z}\xspace}


\def\order{{\ensuremath{\mathcal O}}\xspace}
\def\L{{\ensuremath{\mathcal L}}\xspace}
\def\calL{{\ensuremath{\mathcal L}}\xspace}
\def\calS{{\ensuremath{\mathcal S}}\xspace}
\def\calA{{\ensuremath{\mathcal A}}\xspace}
\def\calD{{\ensuremath{\mathcal D}}\xspace}
\def\calR{{\ensuremath{\mathcal R}}\xspace}

\def\ra                 {\ensuremath{\rightarrow}\xspace}
\def\to                 {\ensuremath{\rightarrow}\xspace}

\newcommand{\stat}{\ensuremath{\text{(stat)}}\xspace}
\newcommand{\syst}{\ensuremath{\text{(syst)}}\xspace}

\newcommand{\sstat}{\ensuremath{\sigma_{\text{stat}}}\xspace}
\newcommand{\ssyst}{\ensuremath{\sigma_{\text{syst}}}\xspace}

\def\pep2{PEP-II}
\def\BF{$B$ Factory}
\def\abf {asymmetric \BF}
\def\sx    {\ensuremath{\sigma_x}\xspace}
\def\sy    {\ensuremath{\sigma_y}\xspace}
\def\sz    {\ensuremath{\sigma_z}\xspace}

\newcommand{\inverse}{\ensuremath{^{-1}}\xspace}
\newcommand{\dedx}{\ensuremath{\text{d}\hspace{-0.1em}E/\text{d}x}\xspace}
\newcommand{\chisq}{\ensuremath{\chi^2}\xspace}
\newcommand{\delm}{\ensuremath{m_{\dstr}-m_{\dz}}\xspace}
\newcommand{\lum} {\ensuremath{\mathcal{L}}\xspace}

\def\gsim{{~\raise.15em\hbox{$>$}\kern-.85em
          \lower.35em\hbox{$\sim$}~}\xspace}
\def\lsim{{~\raise.15em\hbox{$<$}\kern-.85em
          \lower.35em\hbox{$\sim$}~}\xspace}

\def\qsq                {\ensuremath{q^2}\xspace}

\def\kbytes     {\ensuremath{{\rm \,kbytes}}\xspace}
\def\kbsps      {\ensuremath{{\rm \,kbytes/s}}\xspace}
\def\kbits      {\ensuremath{{\rm \,kbits}}\xspace}
\def\kbsps      {\ensuremath{{\rm \,kbits/s}}\xspace}
\def\mbsps      {\ensuremath{{\rm \,Mbits/s}}\xspace}
\def\mbytes     {\ensuremath{{\rm \,Mbytes}}\xspace}
\def\mbps       {\ensuremath{{\rm \,Mbyte/s}}\xspace}
\def\mbsps      {\ensuremath{{\rm \,Mbytes/s}}\xspace}
\def\gbsps      {\ensuremath{{\rm \,Gbits/s}}\xspace}
\def\gbytes     {\ensuremath{{\rm \,Gbytes}}\xspace}
\def\gbsps      {\ensuremath{{\rm \,Gbytes/s}}\xspace}
\def\tbytes     {\ensuremath{{\rm \,Tbytes}}\xspace}
\def\tbpy       {\ensuremath{{\rm \,Tbytes/yr}}\xspace}
%

\def\tb         {\ensuremath{\tan\beta}\xspace}


\newcommand{\as}{\ensuremath{\alpha_{\scriptscriptstyle S}}\xspace}
\newcommand{\asp}{\ensuremath{{\alpha_{\scriptscriptstyle S}\over\pi}}\xspace}
\newcommand{\MSb}{\ensuremath{\overline{\text{MS}}}\xspace}
\newcommand{\LMSb}{%
  \ensuremath{\Lambda_{\overline{\scriptscriptstyle\text{MS}}}}\xspace
}


\newcommand{\tw}{\ensuremath{\theta_{\scriptscriptstyle W}}\xspace}
\newcommand{\twb}{%
  \ensuremath{\overline{\theta}_{\scriptscriptstyle W}}\xspace
}
\newcommand{\Afb}[1]{{\ensuremath{A_{\scriptscriptstyle FB}^{#1}}}\xspace}
\newcommand{\gv}[1]{{\ensuremath{g_{\scriptscriptstyle V}^{#1}}}\xspace}
\newcommand{\ga}[1]{{\ensuremath{g_{\scriptscriptstyle A}^{#1}}}\xspace}
\newcommand{\gvb}[1]{{\ensuremath{\overline{g}_{\scriptscriptstyle V}^{#1}}}\xspace}
\newcommand{\gab}[1]{{\ensuremath{\overline{g}_{\scriptscriptstyle A}^{#1}}}\xspace}


\def\eps{\varepsilon\xspace}
\def\epsK{\varepsilon_K\xspace}
\def\epsB{\varepsilon_B\xspace}
\def\epsp{\varepsilon^\prime_K\xspace}

\def\CP                {\ensuremath{C\!P}\xspace}
\def\CPT               {\ensuremath{C\!PT}\xspace} 

\def\epstag  {\ensuremath{\varepsilon_{\rm tag}}\xspace}
\def\tagfac  {\ensuremath{\epstag(1-2w)^2}\xspace}

\def\rhobar {\ensuremath{\overline \rho}\xspace}
\def\etabar {\ensuremath{\overline \eta}\xspace}
\def\meas {\ensuremath{|V_{cb}|, |\frac{V_{ub}}{V_{cb}}|,
|\varepsilon_K|, \Delta m_{B_d}}\xspace}

\def\Vud  {\ensuremath{|V_{ud}|}\xspace}
\def\Vcd  {\ensuremath{|V_{cd}|}\xspace}
\def\Vtd  {\ensuremath{|V_{td}|}\xspace}
\def\Vus  {\ensuremath{|V_{us}|}\xspace}
\def\Vcs  {\ensuremath{|V_{cs}|}\xspace}
\def\Vts  {\ensuremath{|V_{ts}|}\xspace}
\def\Vtd  {\ensuremath{|V_{td}|}\xspace}
\def\Vub  {\ensuremath{|V_{ub}|}\xspace}
\def\Vcb  {\ensuremath{|V_{cb}|}\xspace}
\def\Vtb  {\ensuremath{|V_{tb}|}\xspace}


\def\stwoa{\ensuremath{\sin\! 2 \alpha  }\xspace}
\def\stwob{\ensuremath{\sin\! 2 \beta   }\xspace}
\def\stwog{\ensuremath{\sin\! 2 \gamma  }\xspace}
\def\mistag{\ensuremath{w}\xspace}
\def\dilution{\ensuremath{\mathcal D}\xspace}
\def\deltaz{\ensuremath{{\rm \Delta}z}\xspace}
\def\deltat{\ensuremath{{\rm \Delta}t}\xspace}
\def\deltamd{\ensuremath{{\rm \Delta}m_d}\xspace}

\newcommand{\fufs}{\ensuremath{f_u/f_s}}\xspace
\newcommand{\fsfu}{\ensuremath{f_s/f_u}}\xspace
\newcommand{\fdfu}{\ensuremath{f_d/f_u}}\xspace
\newcommand{\fsfd}{\ensuremath{f_s/f_d}}\xspace
\newcommand{\fsubd}{\ensuremath{f_D}}\xspace
\newcommand{\fds}{\ensuremath{f_{D_s}}\xspace}
\newcommand{\fsubb}{\ensuremath{f_B}\xspace}
\newcommand{\fbd}{\ensuremath{f_{B_d}}\xspace}
\newcommand{\fbs}{\ensuremath{f_{B_s}}\xspace}
\newcommand{\bsubb}{\ensuremath{B_B}\xspace}
\newcommand{\bbd}{\ensuremath{B_{B_d}}\xspace}
\newcommand{\bbs}{\ensuremath{B_{B_s}}\xspace}
\newcommand{\rgbb}{\ensuremath{\hat{B}_B}\xspace}
\newcommand{\rgbbd}{\ensuremath{\hat{B}_{B_d}}\xspace}
\newcommand{\rgbbs}{\ensuremath{\hat{B}_{B_s}}\xspace}
\newcommand{\rgbk}{\ensuremath{\hat{B}_K}\xspace}
\newcommand{\lqcd}{\ensuremath{\Lambda_{\text{QCD}}}\xspace}

\newcommand{\secref}[1]{Section~\ref{sec:#1}}
\newcommand{\subsecref}[1]{Section~\ref{subsec:#1}}
\newcommand{\figref}[1]{Figure~\ref{fig:#1}}
\newcommand{\tabref}[1]{Table~\ref{tab:#1}}


\newcommand{\epjBase}        {Eur.\ Phys.\ Jour.\xspace}
\newcommand{\jprlBase}       {Phys.\ Rev.\ Lett.\xspace}
\newcommand{\jprBase}        {Phys.\ Rev.\xspace}
\newcommand{\jplBase}        {Phys.\ Lett.\xspace}
\newcommand{\nimBaseA}       {Nucl.\ Instr.\ Meth.\xspace}
\newcommand{\nimBaseB}       {Nucl.\ Instr.\ and Meth.\xspace}
\newcommand{\nimBaseC}       {Nucl.\ Instr.\ and Methods\xspace}
\newcommand{\nimBaseD}       {Nucl.\ Instrum.\ Methods\xspace}
\newcommand{\npBase}         {Nucl.\ Phys.\xspace}
\newcommand{\zpBase}         {Z.\ Phys.\xspace}

\newcommand{\apas}      [1]  {{Acta Phys.\ Austr.\ Suppl.\ {\bf #1}}}
\newcommand{\app}       [1]  {{Acta Phys.\ Polon.\ {\bf #1}}}
\newcommand{\ace}       [1]  {{Adv.\ Cry.\ Eng.\ {\bf #1}}}
\newcommand{\anp}       [1]  {{Adv.\ Nucl.\ Phys.\ {\bf #1}}}
\newcommand{\annp}      [1]  {{Ann.\ Phys.\ {\bf #1}}}
\newcommand{\araa}      [1]  {{Ann.\ Rev.\ Astr.\ Ap.\ {\bf #1}}}
\newcommand{\arnps}     [1]  {{Ann.\ Rev.\ Nucl.\ Part.\ Sci.\ {\bf #1}}}
\newcommand{\arns}      [1]  {{Ann.\ Rev.\ Nucl.\ Sci.\ {\bf #1}}}
\newcommand{\appopt}    [1]  {{Appl.\ Opt.\ {\bf #1}}}
\newcommand{\japj}      [1]  {{Astro.\ Phys.\ J.\ {\bf #1}}}
\newcommand{\baps}      [1]  {{Bull.\ Am.\ Phys.\ Soc.\ {\bf #1}}}
\newcommand{\seis}      [1]  {{Bull.\ Seismological Soc.\ of Am.\ {\bf #1}}}
\newcommand{\cmp}       [1]  {{Commun.\ Math.\ Phys.\ {\bf #1}}}
\newcommand{\cnpp}      [1]  {{Comm.\ Nucl.\ Part.\ Phys.\ {\bf #1}}}
\newcommand{\cpc}       [1]  {{Comput.\ Phys.\ Commun.\ {\bf #1}}}
\newcommand{\epj}       [1]  {\epjBase\ {\bf #1}}
\newcommand{\epjc}      [1]  {\epjBase\ C~{\bf #1}}
\newcommand{\fizika}    [1]  {{Fizika~{\bf #1}}}
\newcommand{\fp}        [1]  {{Fortschr.\ Phys.\ {\bf #1}}}
\newcommand{\ited}      [1]  {{IEEE Trans.\ Electron.\ Devices~{\bf #1}}}
\newcommand{\itns}      [1]  {{IEEE Trans.\ Nucl.\ Sci.\ {\bf #1}}}
\newcommand{\ijqe}      [1]  {{IEEE J.\ Quantum Electron.\ {\bf #1}}}
\newcommand{\ijmp}      [1]  {{Int.\ Jour.\ Mod.\ Phys.\ {\bf #1}}}
\newcommand{\ijmpa}     [1]  {{Int.\ J.\ Mod.\ Phys.\ {\bf A{\bf #1}}}}
\newcommand{\jl}        [1]  {{JETP Lett.\ {\bf #1}}}
\newcommand{\jetp}      [1]  {{JETP~{\bf #1}}}
\newcommand{\jpg}       [1]  {{J.\ Phys.\ {\bf G{\bf #1}}}}
\newcommand{\jap}       [1]  {{J.\ Appl.\ Phys.\ {\bf #1}}}
\newcommand{\jmp}       [1]  {{J.\ Math.\ Phys.\ {\bf #1}}}
\newcommand{\jmes}      [1]  {{J.\ Micro.\ Elec.\ Sys.\ {\bf #1}}}
\newcommand{\mpl}       [1]  {{Mod.\ Phys.\ Lett.\ {\bf #1}}}

\newcommand{\nim}       [1]  {\nimBaseC~{\bf #1}}
\newcommand{\nima}      [1]  {\nimBaseC~A~{\bf #1}}

\newcommand{\np}        [1]  {\npBase\ {\bf #1}}
\newcommand{\npb}       [1]  {\npBase\ B~{\bf #1}}
\newcommand{\npps}      [1]  {{Nucl.\ Phys.\ Proc.\ Suppl.\ {\bf #1}}}
\newcommand{\npaps}     [1]  {{Nucl.\ Phys.\ A~Proc.\ Suppl.\ {\bf #1}}}
\newcommand{\npbps}     [1]  {{Nucl.\ Phys.\ B~Proc.\ Suppl.\ {\bf #1}}}

\newcommand{\ncim}      [1]  {{Nuo.\ Cim.\ {\bf #1}}}
\newcommand{\optl}      [1]  {{Opt.\ Lett.\ {\bf #1}}}
\newcommand{\optcom}    [1]  {{Opt.\ Commun.\ {\bf #1}}}
\newcommand{\partacc}   [1]  {{Particle Acclerators~{\bf #1}}}
\newcommand{\pan}       [1]  {{Phys.\ Atom.\ Nuclei~{\bf #1}}}
\newcommand{\pflu}      [1]  {{Physics of Fluids~{\bf #1}}}
\newcommand{\ptoday}    [1]  {{Physics Today~{\bf #1}}}

\newcommand{\jpl}       [1]  {\jplBase\ {\bf #1}}
\newcommand{\plb}       [1]  {\jplBase\ B~{\bf #1}}
\newcommand{\prep}      [1]  {{Phys.\ Rep.\ {\bf #1}}}
\newcommand{\jprl}      [1]  {\jprlBase\ {\bf #1}}
\newcommand{\pr}        [1]  {\jprBase\ {\bf #1}}
\newcommand{\jpra}      [1]  {\jprBase\ A~{\bf #1}}
\newcommand{\jprd}      [1]  {\jprBase\ D~{\bf #1}}
\newcommand{\jpre}      [1]  {\jprBase\ E~{\bf #1}}

\newcommand{\prsl}      [1]  {{Proc.\ Roy.\ Soc.\ Lond.\ {\bf #1}}}
\newcommand{\ppnp}      [1]  {{Prog.\ Part.\ Nucl.\ Phys.\ {\bf #1}}}
\newcommand{\progtp}    [1]  {{Prog.\ Th.\ Phys.\ {\bf #1}}}
\newcommand{\rpp}       [1]  {{Rep.\ Prog.\ Phys.\ {\bf #1}}}
\newcommand{\jrmp}      [1]  {{Rev.\ Mod.\ Phys.\ {\bf #1}}}  
\newcommand{\rsi}       [1]  {{Rev.\ Sci.\ Instr.\ {\bf #1}}}
\newcommand{\sci}       [1]  {{Science~{\bf #1}}}
\newcommand{\sjnp}      [1]  {{Sov.\ J.\ Nucl.\ Phys.\ {\bf #1}}}
\newcommand{\spd}       [1]  {{Sov.\ Phys.\ Dokl.\ {\bf #1}}}
\newcommand{\spu}       [1]  {{Sov.\ Phys.\ Usp.\ {\bf #1}}}
\newcommand{\tmf}       [1]  {{Teor.\ Mat.\ Fiz.\ {\bf #1}}}
\newcommand{\yf}        [1]  {{Yad.\ Fiz.\ {\bf #1}}}
\newcommand{\zp}        [1]  {\zpBase\ {\bf #1}}
\newcommand{\zpc}       [1]  {\zpBase\ C~{\bf #1}}
\newcommand{\zpr}       [1]  {{ZhETF Pis.\ Red.\ {\bf #1}}}

\newcommand{\hepex}     [1]  {hep-ex/{#1}}
\newcommand{\hepph}     [1]  {hep-ph/{#1}}
\newcommand{\hepth}     [1]  {hep-th/{#1}}

\def\aslund     {\mbox{\tt Aslund}\xspace}
\def\bbsim      {\mbox{\tt BBsim}\xspace}
\def\beast      {\mbox{\tt Beast}\xspace}
\def\beget      {\mbox{\tt Beget}\xspace}
\def\Bta        {\mbox{\tt Beta}\xspace}
\def\betakfit   {\mbox{\tt BetaKfit}\xspace}
\def\cornelius  {\mbox{\tt Cornelius}\xspace}
\def\evtgen     {\mbox{\tt EvtGen}\xspace}
\def\euclid     {\mbox{\tt Euclid}\xspace}
\def\fitver     {\mbox{\tt FitVer}\xspace}
\def\fluka      {\mbox{\tt Fluka}\xspace}
\def\fortran    {\mbox{\tt Fortran}\xspace}
\def\gcalor     {\mbox{\tt GCalor}\xspace}
\def\geant      {\mbox{\tt GEANT}\xspace}
\def\gheisha    {\mbox{\tt Gheisha}\xspace}
\def\hemicosm   {\mbox{\tt HemiCosm}\xspace}
\def\hepevt     {\mbox{\tt{/HepEvt/}}\xspace}
\def\jetset74   {\mbox{\tt Jetset \hspace{-0.5em}7.\hspace{-0.2em}4}\xspace}
\def\koralb     {\mbox{\tt KoralB}\xspace}
\def\minuit     {\mbox{\tt Minuit}\xspace}
\def\objegs     {\mbox{\tt Objegs}\xspace}
\def\paw        {\mbox{\tt Paw}\xspace}
\def\root       {\mbox{\tt Root}\xspace}
\def\squaw      {\mbox{\tt Squaw}\xspace}
\def\stdhep     {\mbox{\tt StdHep}\xspace}
\def\trackerr   {\mbox{\tt TrackErr}\xspace}
\def\turtle     {\mbox{\tt Decay~Turtle}\xspace}

\def\ntrig       {{\tt Ntrig}}
\def\vcal        {{\tt Vcal}}
\def\vthrcomp    {{\tt VthrComp}}
\def\caldel      {{\tt CalDel}}
\def\vtrim       {{\tt Vtrim}}
\def\vana        {{\tt Vana}}
\def\vdig        {{\tt Vdig}}
\def\ctrlreg     {{\tt CtrlReg}}
\def\cals        {{\tt CalS}}
\def\Cal         {{\tt Cal}}
\def\vsh         {{\tt Vsh}}
\def\phscale     {{\tt phscale}}
\def\phoffset    {{\tt phoffset}}
\def\level       {{\tt level}}
\def\clk         {{\tt CLK}}
\def\sda         {{\tt SDA}}

\def\bfbsmmsm  {\ensuremath{(3.66\pm0.14)\times 10^{-9}}}
\def\bfbdmmsm  {\ensuremath{(1.03\pm0.05)\times 10^{-10}}}

\newcommand{\resObsBFBsmm}{\ensuremath{[2.9\pm{0.7}\,(\text{exp})\pm{0.2}\,(\text{frag}) ] \times10^{-9}}}
\newcommand{\resObsBFBsmmLong}{\ensuremath{[2.9\pm{0.6}\stat \pm{0.3}\syst\pm{0.2}\,(\text{frag}) ] \times10^{-9}}}
\newcommand{\resObsBFBdmm}{\ensuremath{(0.8\,^{+1.4}_{-1.3})\times10^{-10}}}
\newcommand{\resObsTauBsmm}{\ensuremath{1.70\,^{+0.61}_{-0.44}}}
\newcommand{\resObsTauBsmmLong}{\ensuremath{[1.70\,^{+0.60}_{-0.43}\stat\pm{0.09}\syst]}}
\newcommand{\sigObsBFBsmm}{\ensuremath{5.6}}
\newcommand{\sigExpBFBsmm}{\ensuremath{6.5}}
\newcommand{\sigObsBFBdmm}{\ensuremath{0.6}}
\newcommand{\sigExpBFBdmm}{\ensuremath{0.8}}
\newcommand{\ulaBFBdmm}{\ensuremath{3.6\times10^{-10}}}
\newcommand{\ulbBFBdmm}{\ensuremath{3.1\times10^{-10}}}
\newcommand{\ulacl}{\ensuremath{95}}
\newcommand{\ulbcl}{\ensuremath{90}}
\newcommand{\ulaExpBdmm}{\ensuremath{3.0\times10^{-10}}}
\newcommand{\ulbExpBdmm}{\ensuremath{2.4\times10^{-10}}}
\newcommand{\resObsTauSplot}{\ensuremath{1.55\ ^{+0.52}_{-0.33}}}
\newcommand{\resObsBFBsmmRunA}{\ensuremath{(2.3\ ^{+1.0}_{-0.8})\times10^{-9}}}
\newcommand{\sigObsBFBsmmRunA }{\ensuremath{3.3}}
\newcommand{\sigExpBFBsmmRunA}{\ensuremath{4.5}}

\markboth{Urs Langenegger}
{Recent results on  \bmm\ decays with the CMS experiment}

\catchline{}{}{}{}{}

\title{Recent results on  \bmm\ decays with the CMS
  experiment\footnote{Based on a seminar given at CERN on Sep.~21, 2019, and published in JHEP, 04, 188 (2020).}}

\author{\footnotesize URS LANGENEGGER}

\address{Paul Scherrer Institute \\CH-5232 Villigen PSI\\
  Switzerland\\
  urs.langenegger@psi.ch\\
  (on behalf of the CMS collaboration)
}

\maketitle

\pub{Received (Day Month Year)}{Revised (Day Month Year)}

\begin{abstract}
Results on \bmm\ decays with the CMS experiment are reported, using
61\invfb\ of data recorded during LHC Run~1 and 2016. With an
improved muon identification algorithm and refined unbinned maximum
likelihood fitting methods, the decay \bsmm\ is observed with a
significance of 5.6 standard deviations. Its branching fraction is
measured to be $\cbfb(\bsmm) = \resObsBFBsmm$, where the first error
is the combined statistical and systematic uncertainty and the second
error quantifies the uncertainty of the \Bs\ and \Bp\ fragmentation
probability ratio. The \bsmm\  effective lifetime is $\tmm =
\resObsTauBsmm\ps$. No evidence for the decay \bdmm\ is found and an
upper limit of $\cbf(\bdmm) < \ulaBFBdmm$ (at 95\%\,confidence level)
is determined. All results are consistent with the standard model of
particle physics.

\keywords{$B$ mesons; leptonic decays; CMS; LHC}
\end{abstract}

\ccode{PACS Nos.: 13.20.He }

\section{Introduction}
\begin{figure}[!htb]
  \centerline{
    \includegraphics[width=0.4\textwidth]{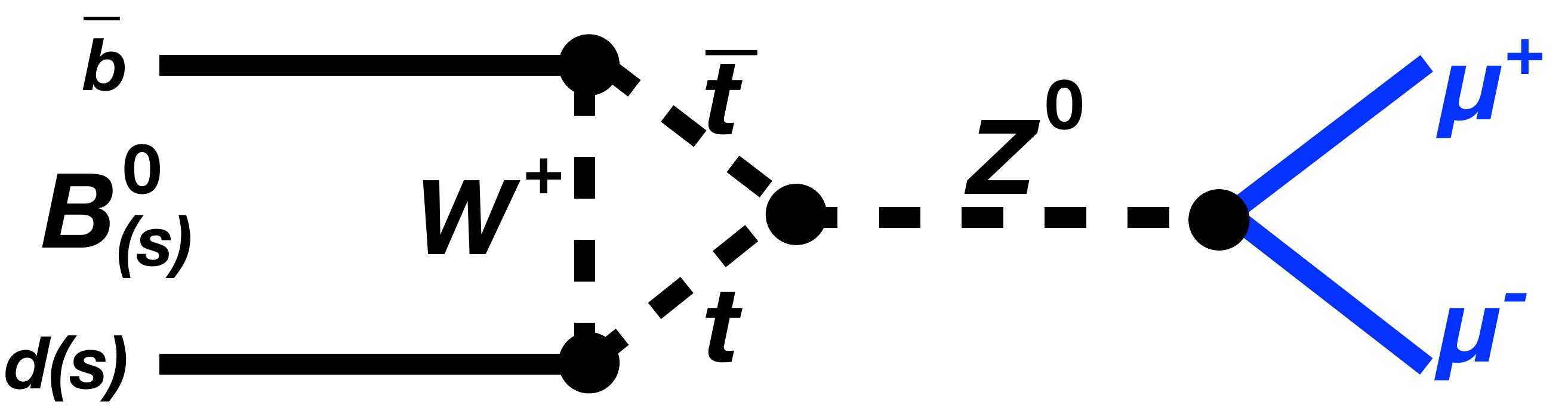}
    \hfill
    \includegraphics[width=0.4\textwidth]{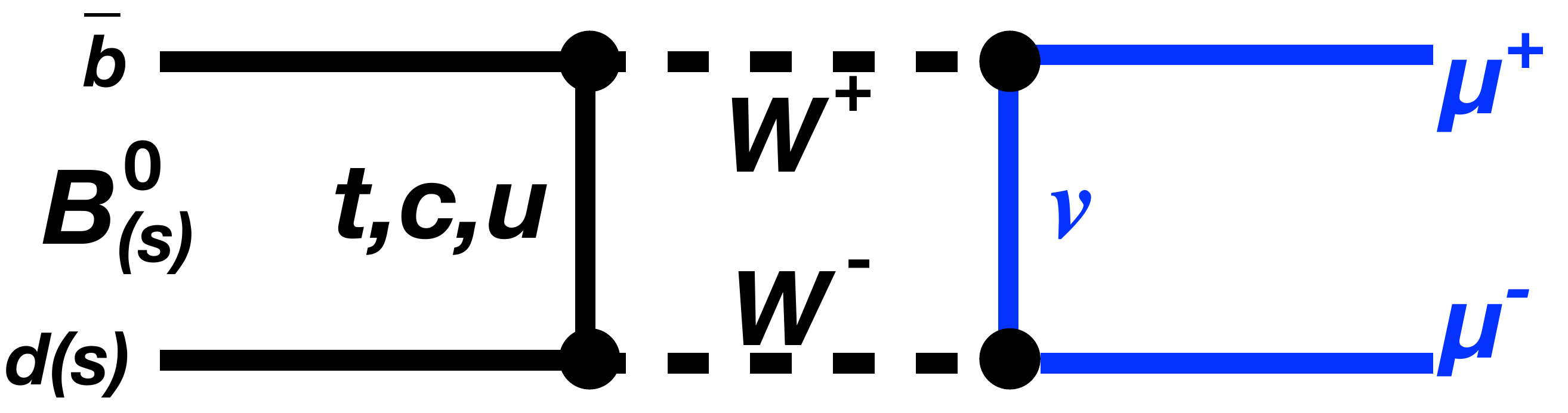}
  }
  \vspace*{8pt}
  \caption{The decays \bsmm\ and
    \bdmm\ are mediated by (left) $Z$-penguin and (right) box
    diagrams. The latter is suppressed by $(m_W/m_t)^2$ with
    respect to the former.\protect\label{f:feynmangraphs}}
\end{figure}

The leptonic $B$ meson\footnote{The symbol $B$ is used to denote \Bz,
  \Bs, and \Bp\ mesons and/or $\Lambda_b$ baryons. Charge conjugation
  is implied throughout, except as noted.} decays \bdmm\ and
\bsmm\ allow precision tests of the standard model (SM) of particle
physics because their branching fractions can be calculated with small
theoretical uncertainties. They are forbidden at tree level in the SM
and are mediated via effective flavor-changing neutral-current
$Z$-penguin and box processes, as illustrated in
Fig.~\ref{f:feynmangraphs}. The branching
fractions~\cite{Bobeth:2013uxa,Hermann:2013kca,Bobeth:2013tba,Beneke:2017vpq,Beneke:2019slt}
in the SM are $\cbf(\bdmm) = \bfbdmmsm$ and $\cbfb(\bsmm) =
\bfbsmmsm$, integrated over the \Bs\ meson decay time. The helicity
suppression of these decays in the SM provides sensitivity to
hypothetical (pseudo-)scalar interactions beyond the SM (BSM).  The
hierarchical nature of the
Cabibbo-Kobayashi-Maskawa\cite{Cabibbo:1963yz,Kobayashi:1973fv} (CKM)
matrix implies that the decay \bdmm\ is CKM-suppressed compared to
\bsmm, since $\vtd < \vts$.

The heavy and light mass eigenstates of the \Bs\ meson, $|\BsLH\rangle
= p|\Bs\rangle \pm q|\Bsb\rangle$ (with $|p|^2 + |q|^2 = 1$), have
different lifetimes. In the absence of \cp\ violation, only the
\cp-odd heavy \Bs\ state, with a lifetime of $\tau_{\BsH} =
1.615\pm0.009\ps$~\cite{pdg2018}, can decay into the dimuon final
state via the SM interactions.
Because this fact is independent of the predicted numerical value of
the branching fraction, it is of high interest to measure also the
\bsmm\ effective lifetime, defined~\cite{DeBruyn:2012wj} as the time
expectation value of the untagged rate by
\begin{equation}
  \tmm \equiv \frac
       {\int_{0}^{\infty} t\,[\Gamma(\Bs(t) \to\mup\mun)+\Gamma(\Bsb(t) \to\mup\mun)]\,dt}
       {\int_{0}^{\infty} [\Gamma(\Bs(t) \to\mup\mun)+\Gamma(\Bsb(t) \to\mup\mun)]\,dt},
\end{equation}
where $t$ is the proper decay time of the \Bs\ meson. Experimentally,
\tmm\ is determined by fitting a single\footnote{In general, the
  untagged decay rate of \Bs\ mesons, with heavy and light
  states, should be described by two exponential functions, not a
  single one.} exponential function, corrected for experimental
artifacts like efficiency and resolution, to the decay time
distribution of \bsmm\ decays.

\begin{figure}[!htb]
  \centerline{
    \hspace*{-1.cm}
    \includegraphics[width=0.56\textwidth]{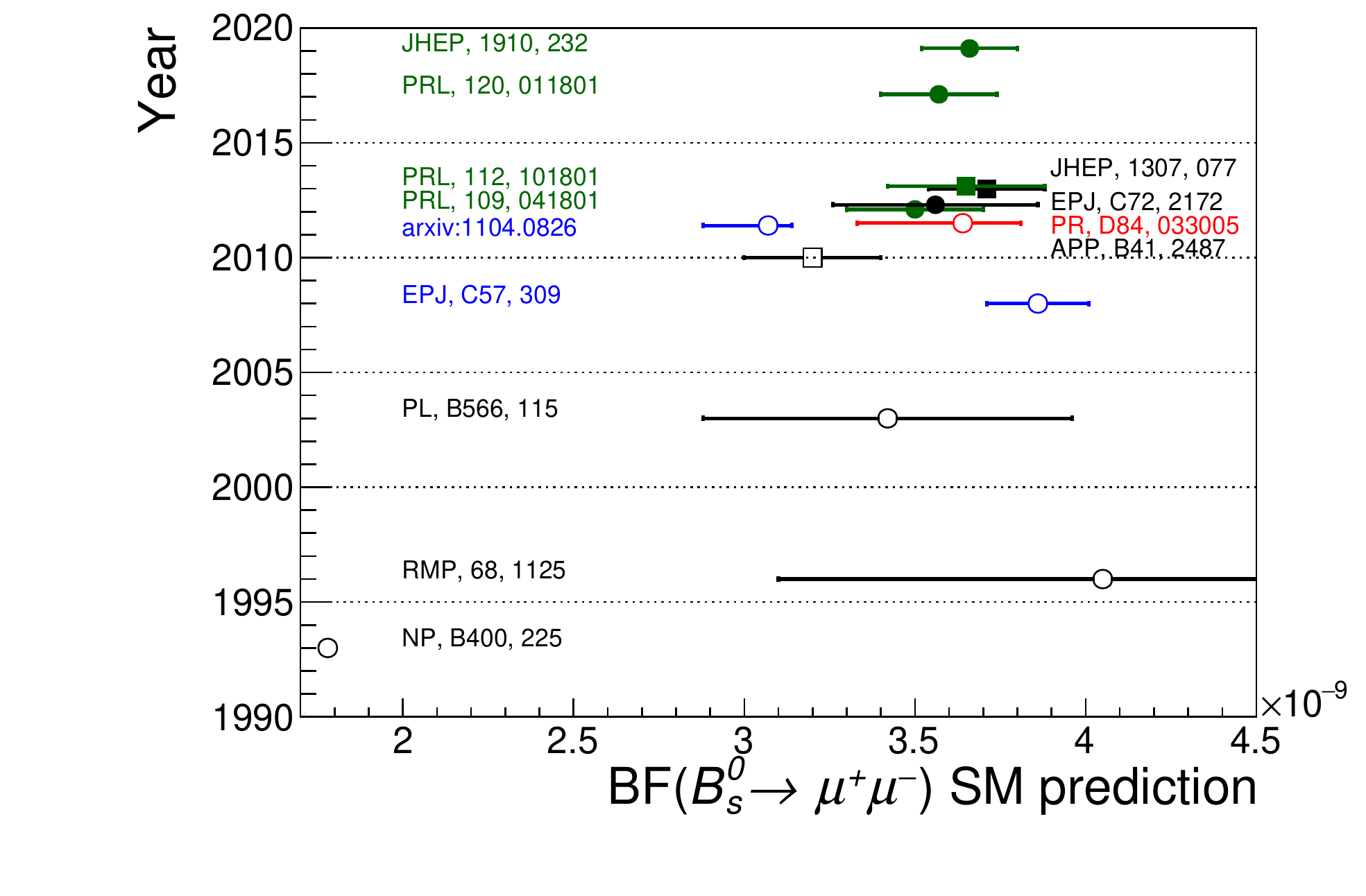}
    \hfill
    \includegraphics[width=0.54\textwidth,trim= 0 -30 0 0 ]{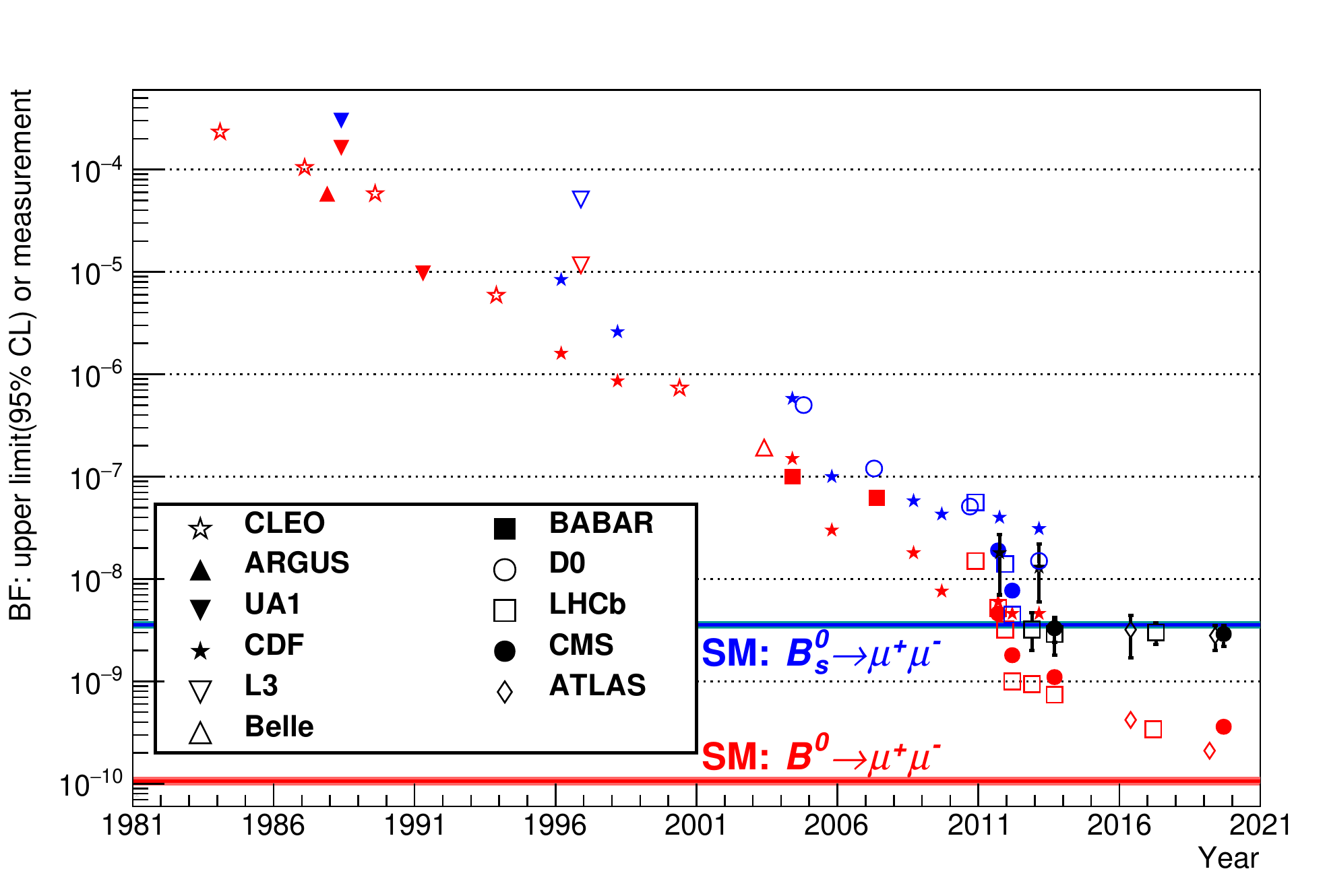}
  }
  \vspace*{8pt}
  \caption{(left) Time evolution of the SM prediction. Note that the
    top quark was not yet discovered at the time of the first
    prediction beyond leading order and that many parameters were
    significantly off from today's values. (right) The experimental
    results for the branching fraction $\cbfb(\bsmm)$, shown with blue
    symbols for upper limits and black symbols for measurements, and
    upper limits for $\cbf(\bdmm)$, shown in red symbols.
    \protect\label{f:evolution}}
\end{figure}

In the SM the branching fractions for these decays have been
calculated beyond leading order in quantum chromodynamics (QCD) since
more than a quarter century (cf.~Fig.~\ref{f:evolution} and references
therein). Most recently, computations of three-loop QCD
corrections~\cite{Hermann:2013kca}, electroweak effects at
next-to-leading order~\cite{Bobeth:2013tba}, and enhanced
electromagnetic corrections~\cite{Beneke:2017vpq,Beneke:2019slt} have
been completed. The (relative) errors are estimated to be smaller than
5\% in the most recent calculation.  By now, the theoretical error
budget is dominated by external parametric uncertainties (either from
CKM matrix element magnitudes or the $B$ meson decay constant,
depending on the number of dynamical quark flavors in the lattice
QCD calculations~\cite{Aoki:2019cca}).

The experimental effort has been pursued both at \ep\en\ machines for
\bdmm\ and at hadron colliders for \bsmm\ (and \bdmm).  The impressive
sensitivity progression in the past four decades is illustrated in
Fig.~\ref{f:evolution} (right). At the Large Hadron Collider (LHC),
the decay \bsmm\ has been measured with at least four standard
deviation ($\sigma$) significances by the ATLAS~\cite{Aaboud:2018mst},
CMS~\cite{Chatrchyan:2013bka}, and LHCb~\cite{Aaij:2013aka}
collaborations. More recently, LHCb~\cite{Aaij:2017vad} and
CMS~\cite{Sirunyan:2019xdu} (with the result discussed here) each have
reached more than $5\sigma$ individually. The \bsmm\ effective
lifetime has been measured first by the LHCb
collaboration~\cite{Aaij:2017vad} and now by the CMS
collaboration~\cite{Sirunyan:2019xdu}. All
confirmed results to date are in agreement with the SM predictions.

\section{Experimental Strategy}

\begin{figure}[!htb]
  \centerline{
    \hspace*{-0.5cm}\includegraphics[width=0.52\textwidth,trim=0 0 0 0]{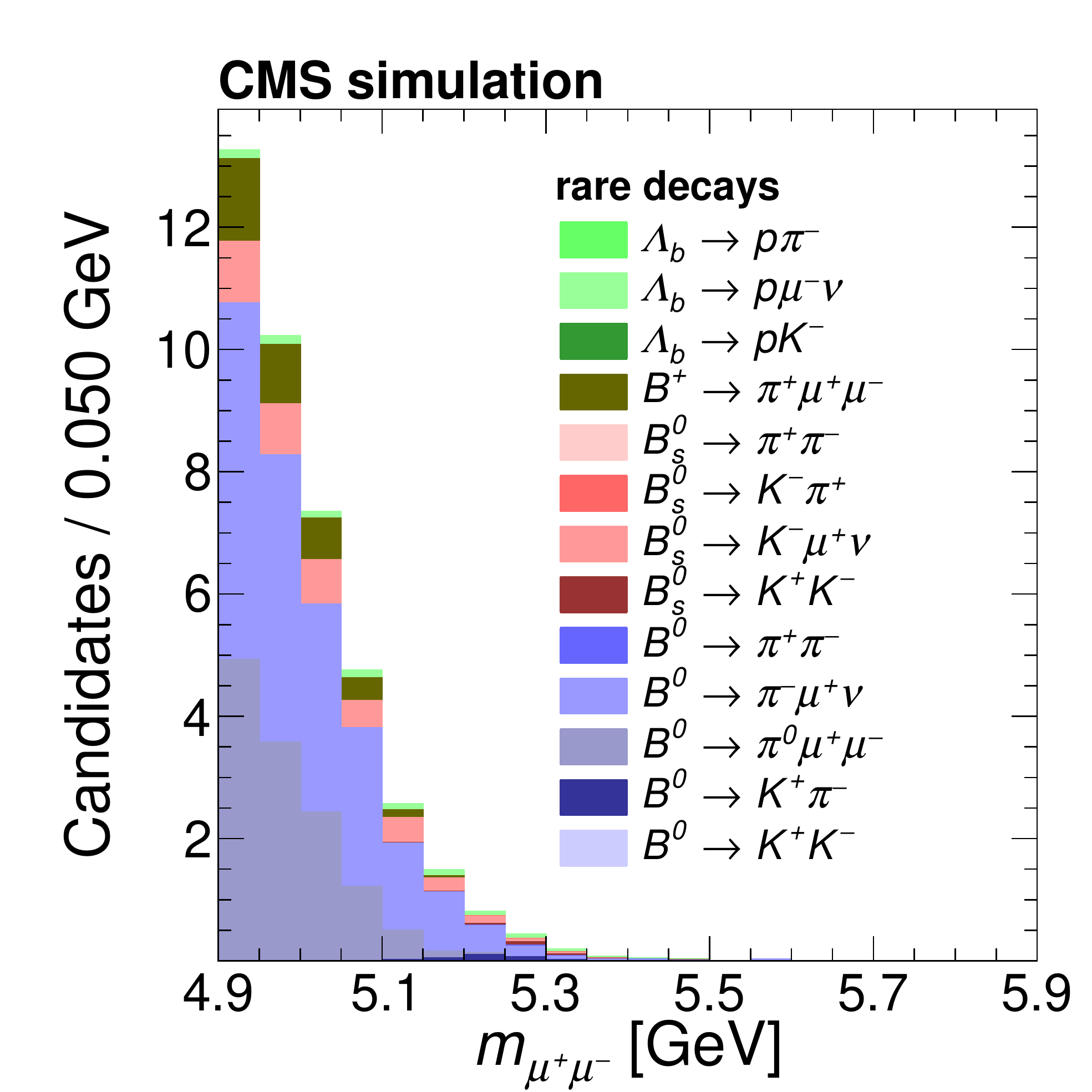}
    \hspace*{0.5cm}\hfill
    \includegraphics[width=0.51\textwidth,trim=0 5 0 0]{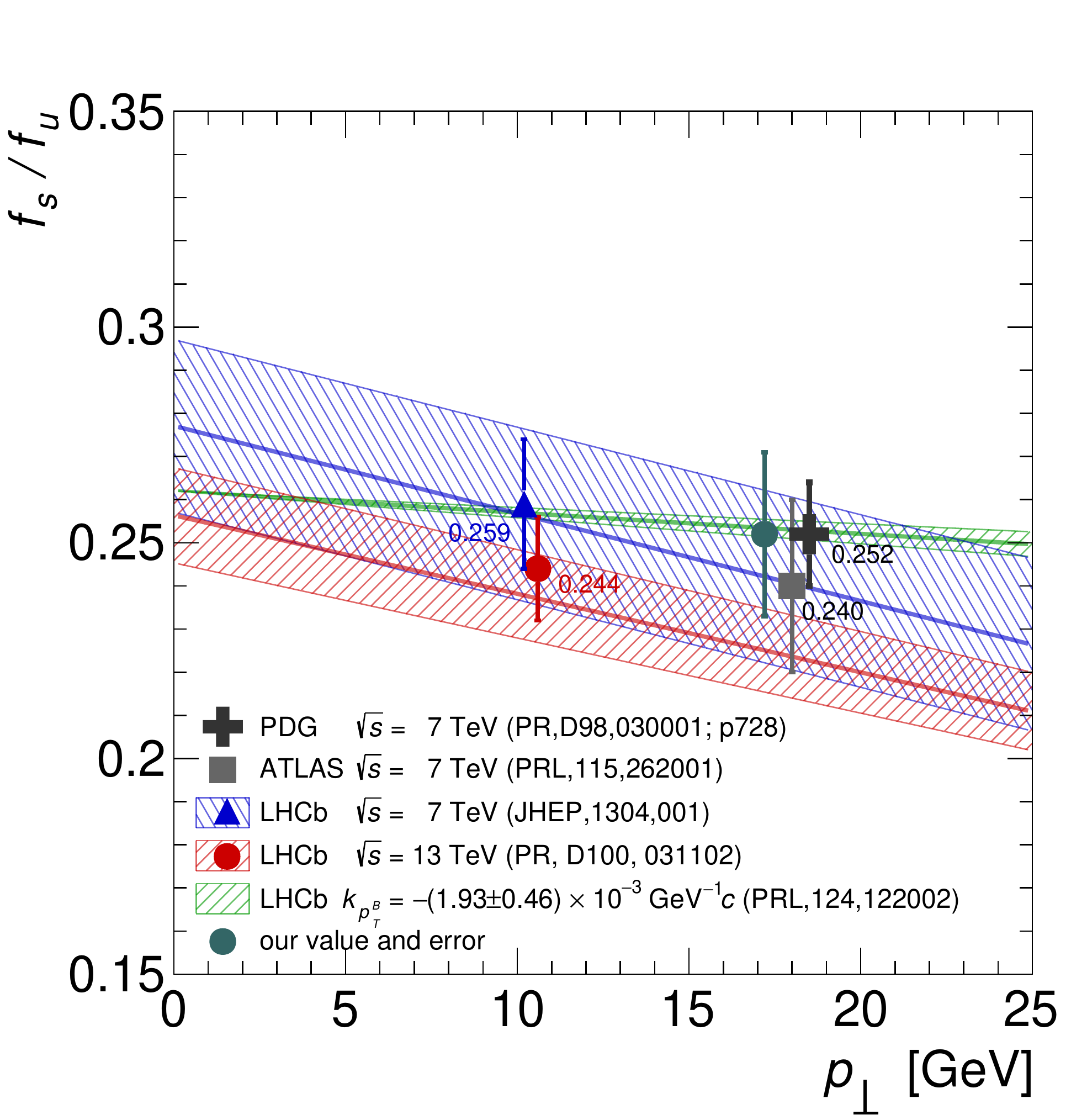}
  }
  \vspace*{8pt}
  \caption{(left) Background invariant mass distribution from rare $B$
    decays with 0, 1, or 2 fake muons (hadrons misidentified as
    muons).  It can be deduced from the plot that these decays are
    much more dangerous for \bdmm\ than for \bsmm\ because of (1)
    their mass distribution covering the \Bz\ mass region ($m_{\Bz} =
    5.280\gev$~\cite{pdg2018}) more strongly than the \Bs\ mass region
    ($m_{\Bs} = 5.367\gev$~\cite{pdg2018}) and (2) the suppressed rate
    of \bdmm\ compared to \bsmm. (right) Illustration of measurements
    of \fsfu, the ratio of the \Bs\ and \Bp\ fragmentation
    probabilities for different mean $B$ meson transverse momentum
    \pt\ and at different center-of-mass energies \sqrts. The bands
    indicate the parametrizations of the LHCb experiment. The value
    adopted by CMS corresponds to the PDG~\cite{pdg2018} value, with
    an ad-hoc enlargement of the error (see text for more details). It
    is interesting to note that the most significant slope
    determination, shown in green, results in the smallest
    slope. \protect\label{f:fsfu}}
\end{figure}

The experimental approach starts with reconstructing dimuon candidates
in a wide invariant mass region. The number of background candidates
is reduced with advanced muon identification algorithms and
multivariate analysis techniques in the selection.  Finally, the
number of signal decays is determined with an unbinned maximum
likelihood fit to the mass distribution and other variables. Signal
\bmm\ decays are characterized by two muons, with an invariant mass
\mll\ around the \Bz\ or \Bs\ mass, originating from a common point in
space where the $B$ meson decayed.  The background has several
components with characteristics that allow its reduction with respect
to the signal:
\begin{itemize}
\item Combinatorial background from two semileptonic $B$ decays or
  from one semileptonic $B$ decay together with a hadron misidentified
  as a muon (fake muon). The two muons do not originate from the same
  point in space. This component is the limiting factor for the
  measurement of \bsmm.
\item Rare decays of a single $B$ hadron, illustrated in
  Fig.~\ref{f:fsfu} (left). They consist of decays with (1) two muons
  (\eg, \bdpimumu, where the pion is not considered in the final state
  reconstruction), (2) one muon combined with a fake muon (\eg,
  \bskmunu), or (3) with two fake muons (\eg, \bskk). These decays
  constitute a dangerous background affecting in particular \bdmm. In
  the first two cases, the mass distribution is leaking into the
  \Bz\ mass region, but the missing particle provides a handle to
  reduce this contribution.  The last case constitutes a
  peaking background near the \Bz\ mass region. The wrong mass
  hypothesis, muon instead of kaon or pion, shifts the mass
  distribution from the \Bs\ mass to lower values.
\end{itemize}
\noindent In the discussion above, a fake muon is a hadron
misidentified in the detector as a muon, either because of its
decay-in-flight or punch-through (hits in the muon system associated
to a charged track).

The branching fraction $\cbf(\bsmm)$ is determined relative to a
normalization sample with a well-known branching fraction. Starting
from Ref.~\refcite{Abazov:2004dj}, the decay \bupsikp, with
$\jpsi\to\mup\mun$, has been used as normalization schematically as
follows:

\begin{equation}
  \cbfb(\bsmm)
  = \frac{N_{\Bs}}{N_{\Bp}} \,
  \frac{f_{u}}{f_{\s}} \,
  \frac{\varepsilon^{\text{tot}}_{\Bp}}{\varepsilon^{\text{tot}}_{\Bs}} \,
  \cbf(\Bp\to\jpsi\Kp)\, \cbf(\jpsi\to\mup\mun)\label{eq:schema},
\end{equation}
where $N_{\Bs} (N_{\Bp})$ is the number of signal \bsmm\ (\bupsikp)
decays, $\varepsilon^\text{tot}_{\Bs}$
($\varepsilon^{\text{tot}}_{\Bp}$) is the total signal (\Bp)
efficiency, $\cbf(\Bp\to\jpsi\Kp) = (1.01\pm0.03)\times10^{-3}$ and
$\cbf(\jpsi\to\mup\mun) = (5.96\pm0.03)\times10^{-2}$~\cite{pdg2018},
and $\fufs$ is the ratio of the \Bp\ and \Bs\ fragmentation
functions. A similar approach is used for $\cbf(\bdmm)$, using $\fdfu
= 1$~\cite{pdg2018}.

In addition to serving as a `normalization sample', the
\bupsikp\ candidates also constitute a large sample of \Bp\ mesons
where the Monte Carlo (MC) simulation can be validated against data.
To compare \Bs\ mesons in data and MC simulation, a `control sample'
of \bspsiphi\ (with $\jpsi\to\mup\mun$ and $\phi\to\Kp\Km$) candidates
is used. Because the analysis relies on MC simulation for the efficiency
determination, the validation of the MC simulation is
essential. Furthermore, these \bpsiX\ samples can be used to study
differences between data and MC simulation regarding $b$ quark
production processes (\eg, by combining such decays with another muon
from a semileptonic $\Bb$ decay) and to study in detail the
$b$ quark hadronization into \Bs\ or \Bp\ mesons.

The usage of \bupsikp\ as a normalization sample is motivated by the
minimal difference in the final state with respect to the signal decay
(one additional charged particle). This leads to a substantial
reduction of the systematic uncertainties. However, it implies a
direct dependence of $\cbfb(\bsmm)$ on \fsfu\ [though not for the
  effective \bsmm\ lifetime or $\cbf(\bdmm)$]. In Fig.~\ref{f:fsfu}
(right) the experimental situation of \fsfu\ measurements and
combinations is illustrated. The LHCb
collaboration~\cite{Aaij:2019eej} obtains a significant slope vs.~the
transverse momentum (\pt) of the $B$ meson, while the ATLAS
collaboration~\cite{Aad:2015cda} and the CMS collaboration (internal
study performed for the result discussed here) see no such
effect. Nevertheless the CMS experiment decided to account for a
hypothetical \pt\ dependence and possible center-of-mass energy
($\sqrts$) dependence by adding an ad-hoc error to \fsfu.  An
uncertainty of 0.008 is derived from the difference between the value
of \fsfu\ in Ref.~\refcite{pdg2018}, obtained at $\sqrt{s}=7\tev$, and
that in Ref.~\refcite{Aaij:2019pqz}, obtained at $\sqrt{s}=13\tev$.
In addition, with the parametrization of the \pt\ dependence in
Ref.~\refcite{Aaij:2019pqz}, a difference of 0.013 is determined
between the \fsfu\ values at the average \pt\ of
Ref.~\refcite{Aaij:2019pqz} and the average \pt\ of the
\bsmm\ candidates in this analysis. Ref.~\refcite{Aaij:2019eej} would
imply a much smaller \pt\ dependence, but was published too late to be
included. In summary, the CMS experiment uses $\fsfu =
0.252\pm0.012(\text{exp})\pm 0.015(\text{CMS})$ where the first error
is from the PDG~\cite{pdg2018} and the second error is the ad-hoc
error of CMS.

In the future, a normalization to other decay modes may provide a less
contentious solution. While $\cbf(\bspsiphi)$ normally has a
dependence on \fsfd\ (or \fsfu) when determined at hadron colliders,
results obtained at \ep\en\ machines at the \FiveS\ provide additional
input and may lead to a smaller overall error.

To avoid a possible bias, the analyses searching for and measuring the
\bmm\ decays have been pursued as `blind' analyses since a long
time. This implies that a signal region, often defined in terms of the
invariant mass, is hidden during the development and optimization of
the analysis methodology. To take full advantage of analysis
improvements (\eg, improved muon identification, re-processed data
with better tracking resolutions, etc.) it is advantageous to
re-analyze the old, previously published, datasets. This implies a
`re-blinding' of the data, which is, however, not a problem because
the improvements normally change the set of candidates noticeably.  An
alternative would be to combine new results only statistically with
the old results, albeit at a loss of sensitivity. In this analysis,
the first approach is used and the dimuon mass range $5.2 < \mll <
5.45\gev$ was kept (re-)blinded until the entire selection and fitting
procedure was finalized.

\section{Detector and Data}
The data for this analysis was collected by the CMS
experiment~\cite{Chatrchyan:2008zzk} in LHC Run~1 (25\invfb) and in
2016 (36\invfb), as summarized in Table~\ref{t:channels}. The CMS
experiment is very well suited for \bmm\ measurements because its
silicon tracker, composed of a pixel detector with 66 million pixels
of $100\mum\times150\mum$ and a micro-strip detector with 10 million
strips with pitches between $80$ and $180\mum$, provides outstanding
three-dimensional (3D) vertexing and tracking capabilities in a very
homogeneous solenoidal magnetic field of $3.8\Tesla$. The tracker is
divided into barrel and endcap parts. In Run 2, the micro-strip
detector was subject to operational instabilities and the data are
therefore divided into two separate data-taking periods, 2016A and
2016B, of 16\invfb and 20\invfb, respectively. The systematic error of
the tracking efficiency is
estimated~\cite{Khachatryan:2010pw,CMS-DP-2018-050} to be 4\% (2.3\%)
in Run 1 (2016).

\begin{table}[h]
  \tbl{Summary of the data, together with center-of-mass energy
    \sqrts, the integrated luminosity \clu, the pileup quantified as
    the average number of $pp$ collision vertices reconstructed as
    primary vertices $\langle N_{\text{PV}}\rangle$, and the channel
    definition based on the pseudorapidity of the most-forward muon
    $|\eta_\mu^{\text{f}}|$. Note that in Run~2 (2016A and 2016B) the
    region $|\eta_\mu^{\text{f}}| > 1.4$ (the forward channel of
    Run~1) is no longer present because of trigger rate constraints.
  }{
    \begin{tabular}{@{}rccccc@{}}
      \toprule
      Data-taking    &\sqrts  &\clu  &Pileup &\multicolumn{2}{c}{Channels}\\
      period    &[\tev] &[\invfb] &$\langle N_{\text{PV}}\rangle$ &central &forward\\
      \toprule
      2011  &7  &5  &8  &$0 < |\eta_\mu^{\text{f}}| < 1.4$  &$1.4 < |\eta_\mu^{\text{f}}| < 2.1$ \\
      2012  &8  &20 &15 &$0 < |\eta_\mu^{\text{f}}| < 1.4$  &$1.4 < |\eta_\mu^{\text{f}}| < 2.1$ \\
      2016A &13 &16 &18 &$0 < |\eta_\mu^{\text{f}}| < 0.7$  &$0.7 < |\eta_\mu^{\text{f}}| < 1.4$ \\
      2016B &13 &20 &18 &$0 < |\eta_\mu^{\text{f}}| < 0.7$  &$0.7 < |\eta_\mu^{\text{f}}| < 1.4$ \\
      \botrule
    \end{tabular}
    \label{t:channels}
  }
\end{table}

Muons are detected in four muon stations, using three complementary
detector types, interspersed among the steel flux-return plates.
Standalone muons are formed from hits in the muon stations and
combined with silicon tracker tracks to form so-called global
muons~\cite{Chatrchyan:2012xi,Sirunyan:2018}. A dedicated boosted
decision tree (BDT) was trained separately for Run 1 and Run 2 data to
obtain the best possible hadron-to-muon misidentification
probability. The starting point for this BDT are global muons. The
variables used in the BDT are based on measurements from the (1)
silicon tracker, (2) the muon system, and (3) the combined global muon
reconstruction. This new muon BDT achieves an average muon
misidentification probability of $6\times10^{-4}$ and $10^{-3}$ for
pions and kaons, respectively, with a muon identification efficiency
of about 75\%. Compared to the previous
analysis~\cite{Chatrchyan:2013bka}, the muon BDT is operated at a
significantly lower muon misidentification rate and roughly the same
muon identification efficiency. The muon BDT is extensively validated
with kinematically identified samples of muons, pions, kaons, and
protons (from the decays $\jpsi\to\mup\mun$, $\KS\to\pip\pim$,
$\phi\to\Kp\Km$, and $\Lambda\to p \pim$, respectively) in data and MC
simulation. All distributions of variables used in the muon BDT, the
BDT discriminator distributions, and the absolute muon
misidentification probability are found to be consistent between data
and simulation. The systematic error for the muon efficiency is
determined from the difference of the efficiency ratio of the muon BDT
discriminator requirement for \bupsikp\ and \bspsiphi\ between data
and simulation, which agrees to better than 3\%. The systematic error
on the muon misidentification is derived from a direct comparison of
the absolute muon misidentification probability in data and simulation
(10\% relative uncertainty for pions and kaons). For protons, the very
small sample size of $\Lambda\to p \pim$, with a misidentified proton,
does not allow this approach and the error of the average muon
misidentification probability, in data and simulation, is used instead
(60\% relative uncertainty).

The trigger~\cite{Khachatryan:2016bia} of the CMS experiment has two
stages: the first stage is based on custom hardware processors and
selects two muons with either no or minimal \ptmu\ threshold (because
of the strong magnetic field, there is an implicit \ptmu\ threshold of
about $3.5\gev$ in the central region). The second stage, the
high-level trigger (HLT) consists of a processor farm running the full
event reconstruction software with reduced sets of calibration
constants. The normalization sample is triggered with a setup that is
very similar to the signal setup with the exception that the two muons
must be consistent with originating from a \jpsi\ meson from a $B$
decay (`displaced \jpsi'). The signal (normalization) trigger
efficiency varies over the data-taking periods from 65--75\%
(50--75\%). The systematic uncertainty on the ratio of the trigger
efficiency is estimated to be 3\%.

The tracking detectors of the CMS experiment induce a strong
pseudorapidity ($\eta$) dependence of the mass resolution. Therefore
the analysis sensitivity benefits from subdividing the data into
`channels', according to the $\eta$ of the most forward muon. Because
of trigger changes over the years, the boundary between the `central'
and `forward' channels is different between Run 1 and Run 2. In total,
there are eight channels in the analysis (central and forward in four
data-taking periods) as summarized in Table~\ref{t:channels}.

At the instantaneous luminosity of the LHC, multiple $pp$ interactions
(pileup) occur in each bunch crossing. Tracks from other $pp$
interaction vertices (primary vertices or PVs) increase the
combinatorial background and complicate the determination of key
variables in the selection, as discussed below. Table~\ref{t:channels}
lists the average number of PVs reconstructed in the different
data-taking periods.

\section{Candidate Selection}
\begin{figure}[!htb]
  \centerline{
    \includegraphics[width=0.5\textwidth]{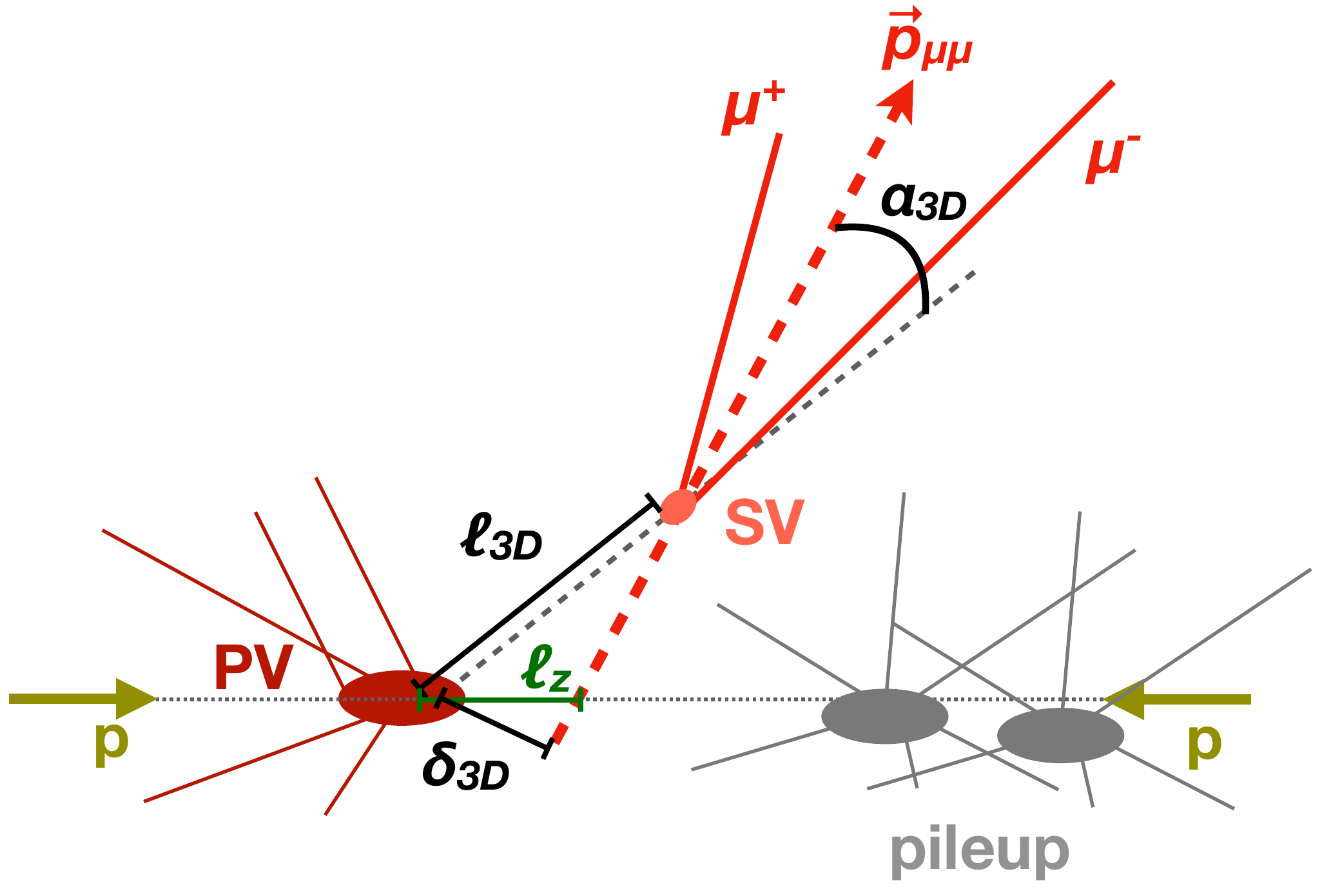}
    \hfill
    \includegraphics[width=0.4\textwidth]{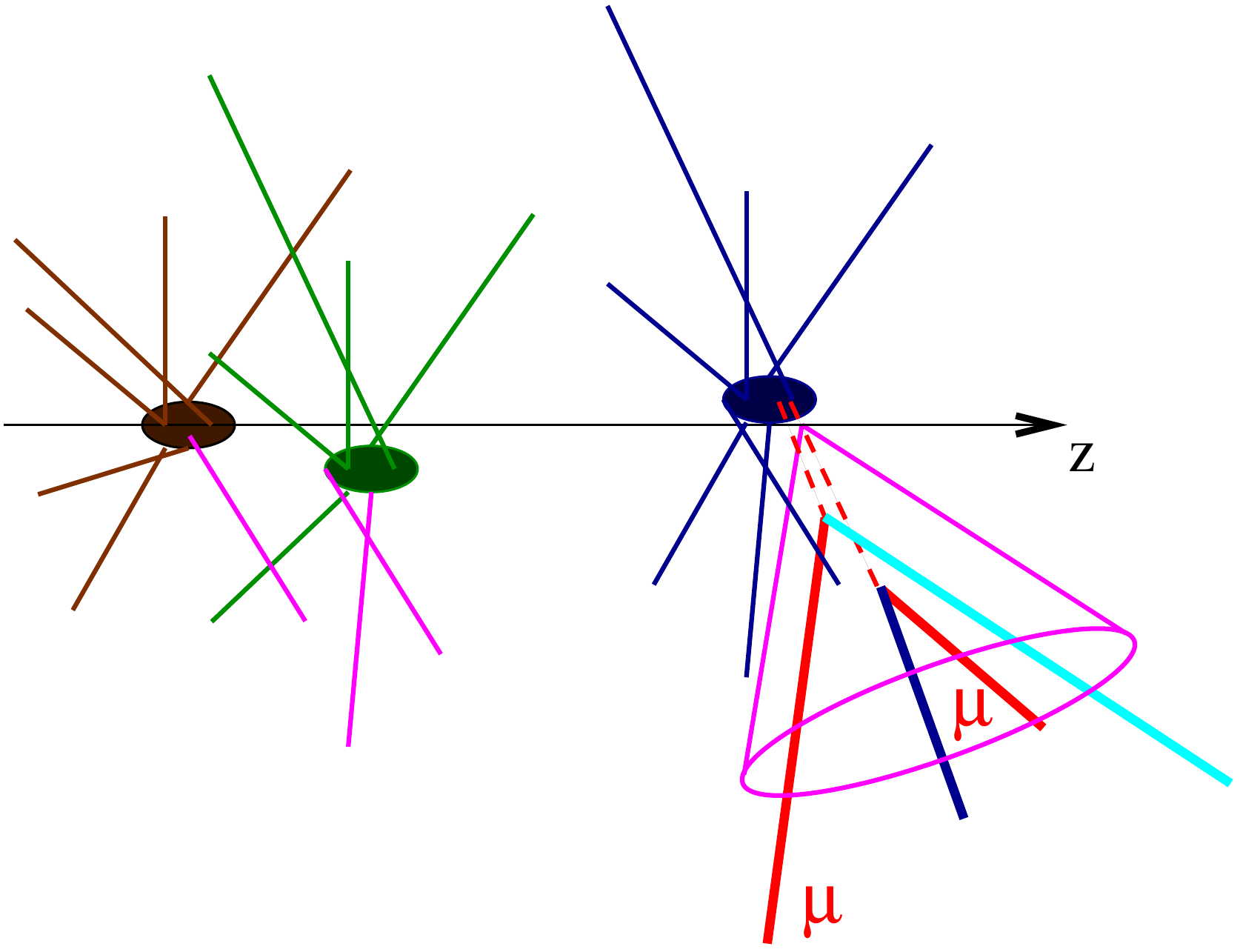}
  }
  \vspace*{8pt}
  \caption{(left) Sketch of a signal decay with the variables involved
    (see text for a description). (right) Sketch of a background decay
    showing aspects of the candidate isolation. Tracks from pileup PVs
    that would fall within the isolation cone (shown in pink) are
    excluded from the isolation calculation by requiring that all
    tracks considered are associated with the $B$-PV or are close to
    the SV (see text for more details).  \protect\label{f:variables}}
\end{figure}

The \bmm\ candidate reconstruction starts with two global muons with
$\ptmu>4\gev$ and a small distance of closest approach $\dca < 0.8\cm$
between their trajectories. The two muons are constrained to originate
from a common point (secondary vertex or SV) and to have an invariant
mass $4.9 < \mll < 5.9\gev$, after refitting their momenta to include
the SV as an additional hit. For each reconstructed $B$ candidate one
specific PV is chosen as the $B$-meson origin (denoted below as
$B$-PV), based on the longitudinal impact parameter $\ell_z$ along the
beam axis of the extrapolated $B$-meson trajectory.  The
PVs are refitted by excluding tracks from the $B$ candidate. The
variables involved in the selection are sketched in
Fig.~\ref{f:variables} (left) for a signal event.  The \bmm\ selection
exploits the differences between signal and background. Signal
\bmm\ decays are characterized by a SV with a good fit \chidof,
separated from the $B$-PV by a large flight length $\fl$ and
significance \fls, where $\sigma(\fl)$ is the error of \fl. The
$B$-meson proper decay time is measured as $t = \mll\fl/\pmm$ in 3D
space.  The $B$-meson momentum is well aligned with the flight
direction (the direction from the $B$-PV to the SV), implying a small
opening angle \pa\ and a small $B$ impact parameter \ip\ and
significance \ips.

For a signal decay, the two muons are the only final-state particles
of the $B$ meson decay, and therefore the $B$ meson and the muons are
isolated, \ie, not many other charged tracks are nearby. This
isolation is quantified by two sets of variables. The first set
determines a macroscopic isolation $I$ using
\begin{equation}
  I = \frac{\ptb}{\ptb + \sum_{\mathrm{trk}}\pttrk},\label{eq:iso}
\end{equation}
\noindent where the sum includes all tracks, not part of the $B$
candidate, inside a cone with radius $\Delta R = \sqrt{(\Delta\eta)^2
  + (\Delta\varphi)^2} < 0.7$, where $\Delta\eta$ $(\Delta\varphi)$ is
the difference between the track and the $B$ candidate $\eta$
(azimuthal angle $\varphi$). Because of tracks from other PVs not
related to the $B$-PV, it is essential to only use tracks that are
either associated to the $B$-PV or are close to the SV. Without these
requirements, the tracks from other pileup PVs would bias $I$, as
indicated in Fig.~\ref{f:variables} (right) for a background
event. The tracks in the sum must fulfill $\pt > 0.9\gev$ and have
$\dcasv< 0.05\cm$ with respect to the SV. These requirements were
optimized with \bpsiX\ decays for strongest background rejection and
best agreement between data and simulation. Similar isolation
variables, \isomuone\ and \isomutwo, are determined for each of the
two muons, although with different track requirements in the sum,
$\pt>0.5\gev$ and $\dcasv<0.1\cm$, and a smaller cone radius of $\Delta
R = 0.5$.

The second set of isolation variables is based on microscopic
observables determined with tracks of $\pt>0.5\gev$ that are not part
of the $B$ candidate and are not associated to any non-$B$-PV (\ie, if
they are associated to a PV it must be the $B$-PV): the minimum
distance of closest approach, \docatrk, of any qualifying track to the
SV and \closetrk, the number of tracks with $\dcasv < 0.03\cm$ to the
SV.

Many of these variables are correlated with each other. For instance,
the flight length and its significance are correlated with \closetrk,
and the isolation variables $I$, \isomuone, and \isomutwo are
correlated with each other. A robust approach in such a situation is
to train a multivariate analysis technique~\cite{Hocker:2007ht} in the
form of a BDT. This selection BDT was trained on \bsmm\ and
\bdmm\ signal decays from MC simulation and combinatorial background
from the data sideband $5.45 < \mll < 5.9\gev$. Since data events are
used in the training, the MC and data samples were randomly split into
three subsets to ensure that the {\it training\/} and {\it validation\/} of a BDT are
performed on subsets completely independent of its {\it application}. This
procedure implies that three BDTs are required for the analysis of
each channel.  A preselection removes candidates with extreme outlier
values in the variables and requires the SV to be well separated from
the $B$-PV, $\fls > 4$. With the subdivision of the data into three
subsets per channel, the number of events available for training is
somewhat limited after the preselection (at least 6000 candidates in
any subset). For the per-channel optimization of the BDT
configuration, a set of core variables [\fls, \pa, \ips, \docatrk,
  \chidof, \closetrk, \iso, \isomuone, \isomutwo] is iteratively
combined with a subset of other $B$-candidate variables [\dca, \ip, \fl, \flsxy,
\ptb, $\etab$]. The best BDT configurations are chosen based on (1) the
maximum of $S/\sqrt{S+B}$, where $S$ ($B$) is the expected
\bsmm\ signal (combinatorial background, extrapolated from the
sideband) yield in the mass region $5.3 < \mll\ < 5.45\gev$ and (2)
the visual assessment of the agreement between data and MC simulation
using large samples of exclusive \bpsiX\ candidates.

\begin{figure}[!htb]
  \centerline{
    \includegraphics[width=0.5\textwidth]{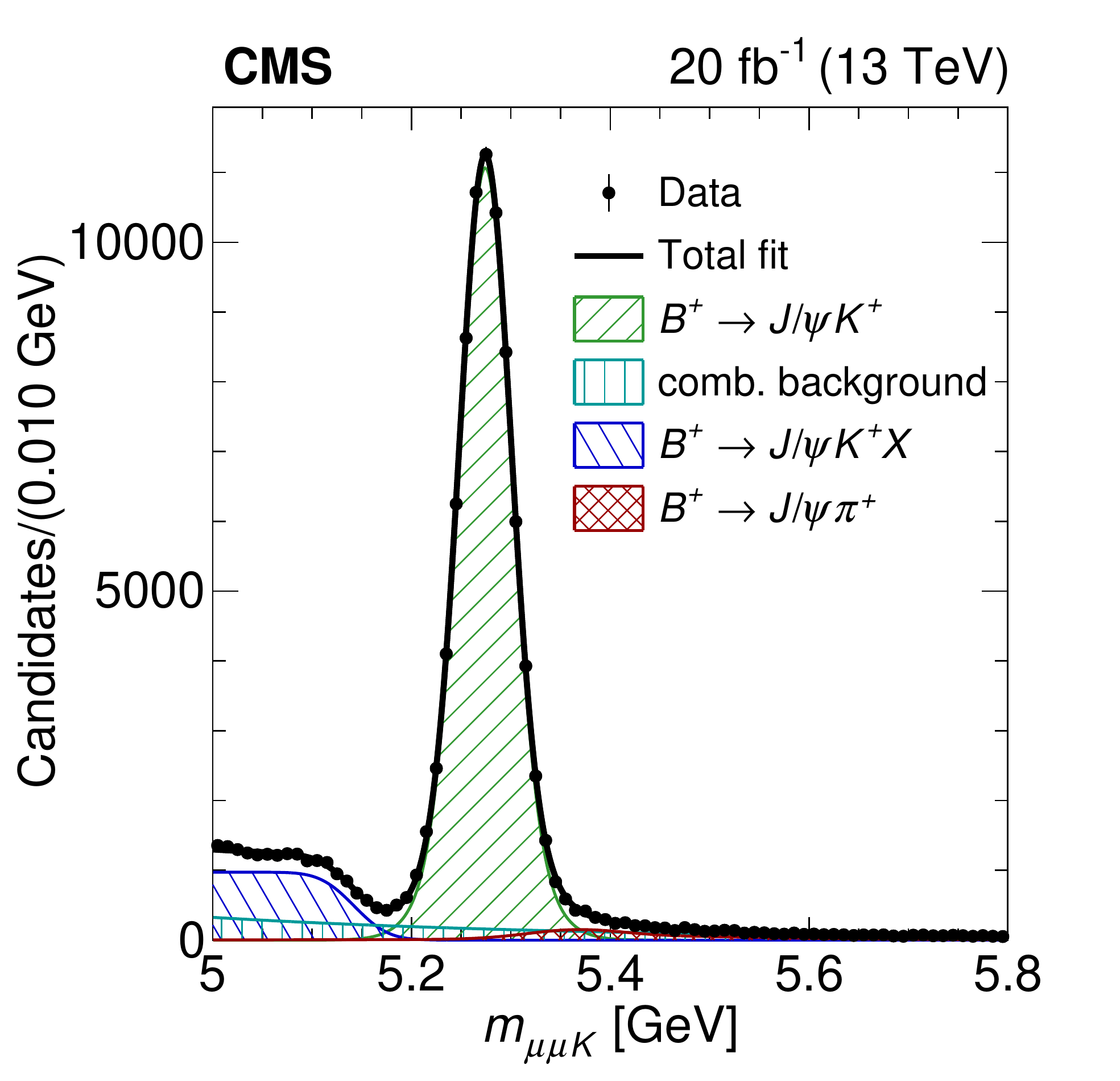}
    \hfill
    \includegraphics[width=0.5\textwidth]{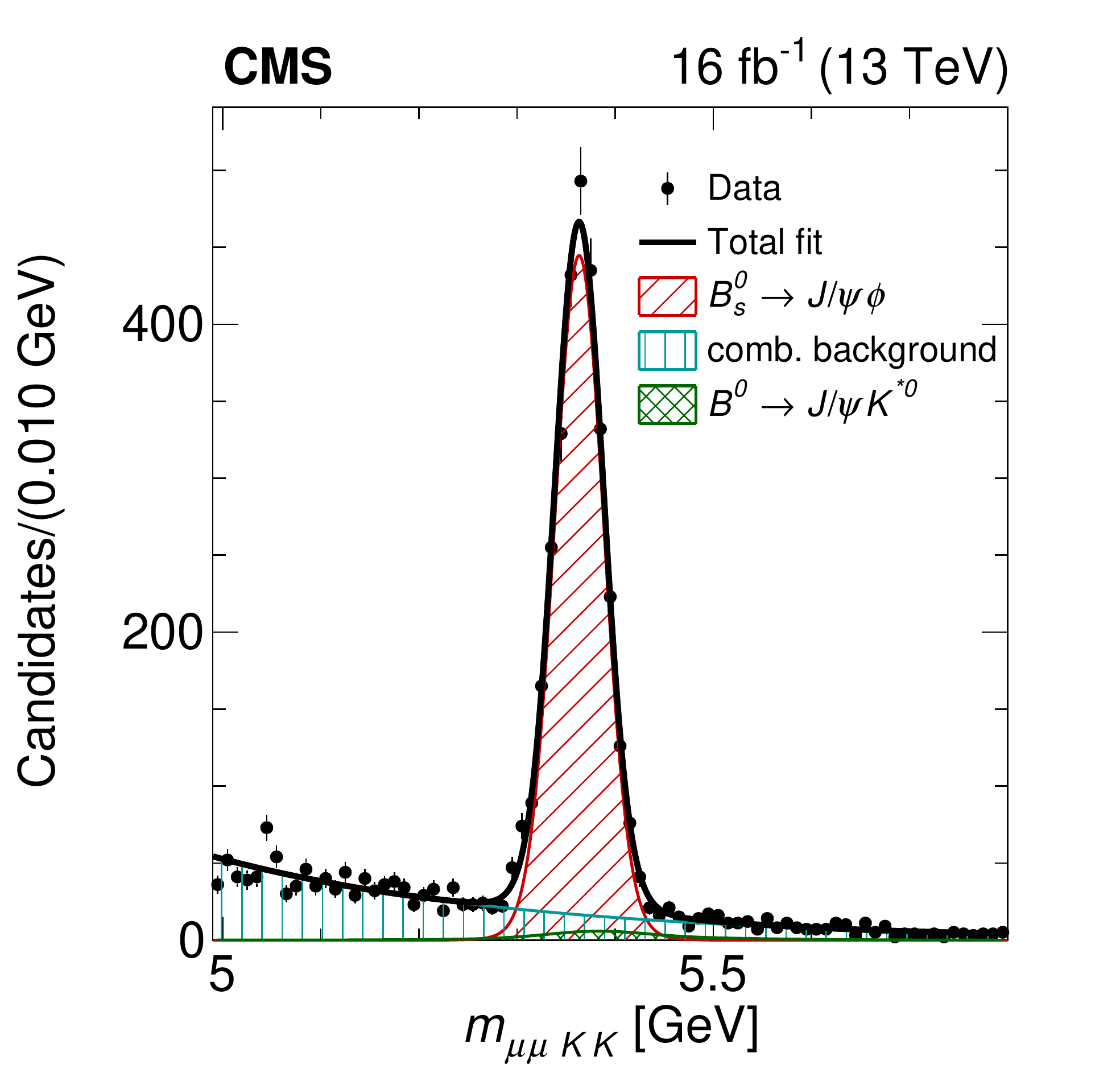}
  }
  \vspace*{8pt}
  \caption{Invariant mass distribution for (left) the normalization
    sample \bupsikp\ and (right) the control sample \bspsiphi. The
    data are shown by solid black circles, the result of the fit is
    overlaid with the thick solid black line, and the different
    components are shown with hatched
    regions. \protect\label{f:normalization}}
\end{figure}

The decays \bupsikp\ and \bspsiphi, both with $\jpsi\to\mup\mun$,
allow the validation of the selection BDT and serve as normalization
and control samples, respectively. Their reconstruction starts with
two oppositely charged muons with $\ptmu>4\gev$, $\ptmm > 7\gev$, and
$2.9 < \mll < 3.2\gev$. They are combined with one or two tracks,
respectively, assumed to be kaons and with $\pt>0.6\gev$. To reduce
combinatorial background, the maximum distance of closest approach
between any pair of tracks is required to fulfill $\maxdoca <
0.08\cm$. For \bspsiphi, the two tracks must fulfill $1.01 <
m_{\Kp\Km} < 1.03\gev$. To allow the selection of the
\bpsiX\ candidates with the same selection BDT as for the signal
\bmm\ decays, their variable distributions should mirror the
corresponding ones from signal decays. This implies that the SV
\chidof\ is not based on the full SV fit with three or four tracks,
but only on the dimuon vertex fit. In addition, for all isolation
variables, the kaon track(s) are not part of the track sums of
Eq.~\ref{eq:iso}. Example mass distributions, together with fits to
the data using signal (double Gaussian functions with a common mean)
and background components, are shown in
Fig.~\ref{f:normalization}. The background components for
\bupsikp\ contain an exponential function for the combinatorial
background, an error function for partially reconstructed
$B\to\jpsi\Kp X$ decays, and a MC simulation based shape for the
peaking background from \bupsipi\ (fixed to 4\%\cite{pdg2018} of the
total signal yield). For \bspsiphi\ the background is parametrized by
an exponential function for the combinatorial component and a MC
simulation based shape for the peaking background from \bdpsikstar,
with $\Kstarz\to\Kp\pim$ (where the pion is treated as a kaon). The
total \bupsikp\ normalization yield used for the determination of
$\cbf(\bsmm)$ is $N_{\Bp} = (1.43\pm0.06)\times10^6$, where the error
is dominated by the systematic component of (relative) 4\%, obtained
from the yield comparison between fits without and with \jpsi-mass
constraints.

\begin{figure}[!htb]
  \centerline{
    \includegraphics[width=0.45\textwidth]{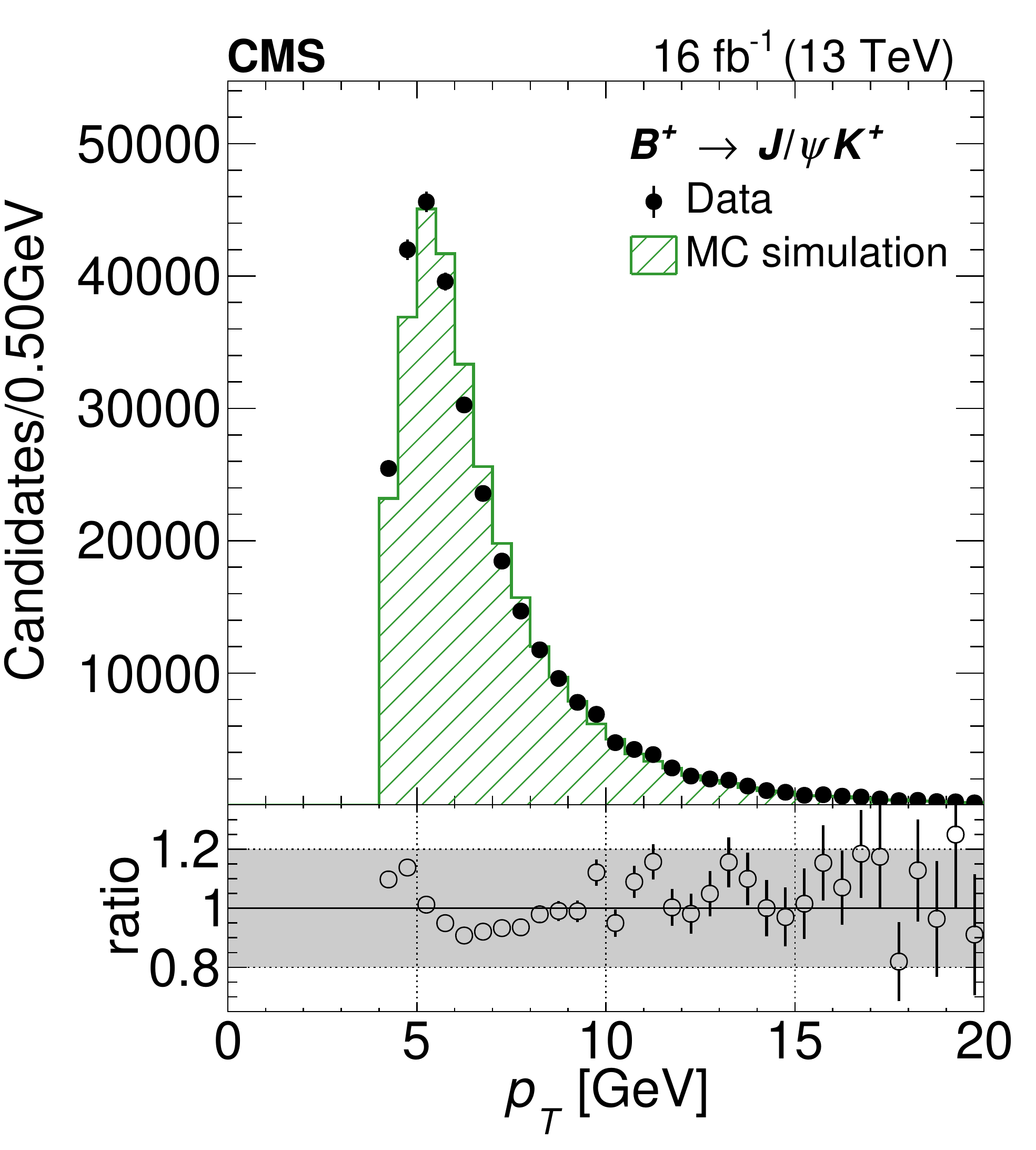}
    \hfill
    \includegraphics[width=0.45\textwidth]{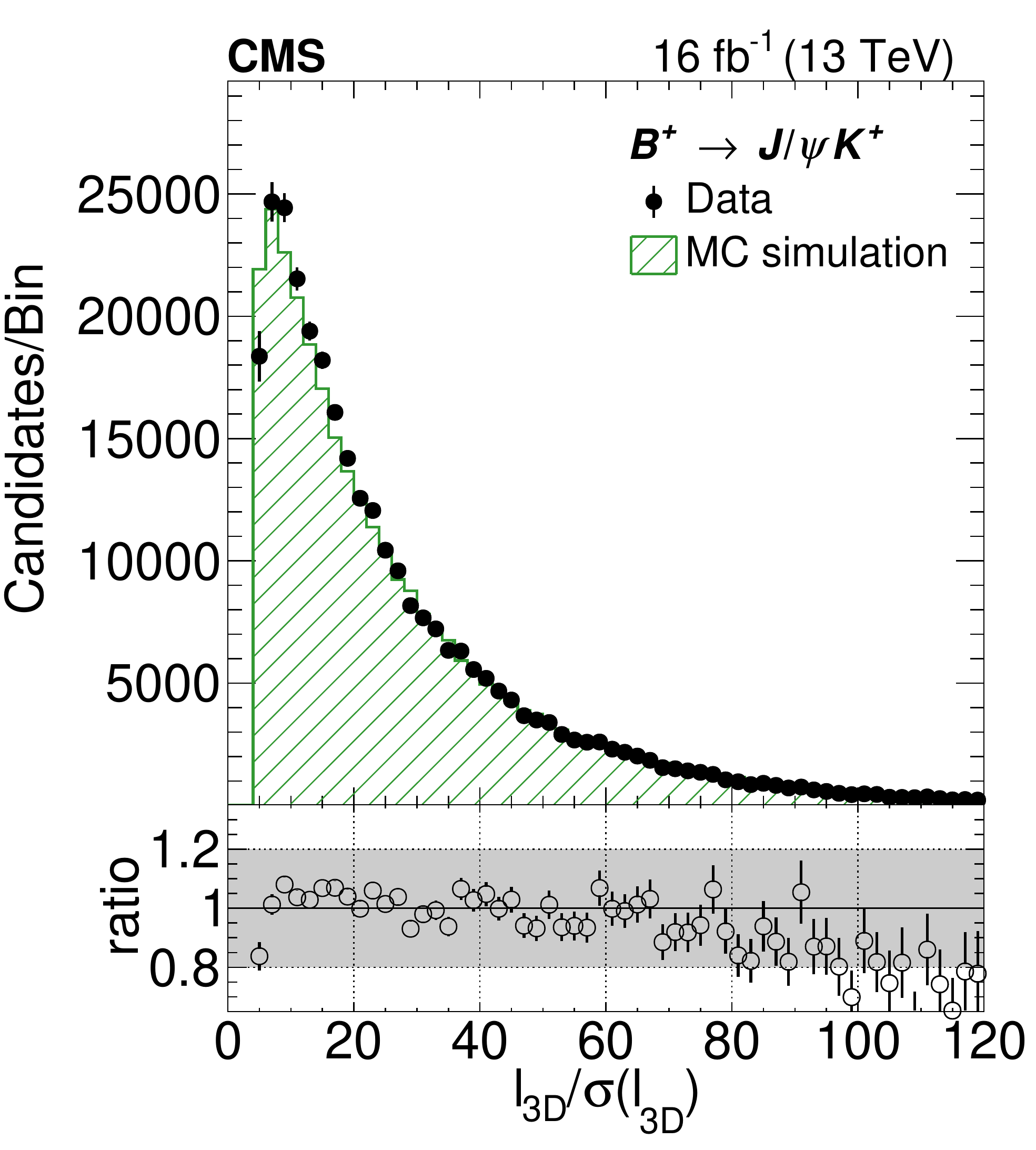}
  }
  \vspace*{8pt}
  \caption{Comparison of the measured and simulated distributions of
    (left) the subleading muon \pt\ and (right) the flight length significance
    \fls\ for \bupsikp\ decays in data and MC
    simulation. Background-subtracted data are shown by solid black
    circles, the MC simulation by hatched histograms. In the lower
    panels the ratio between data and MC simulation is shown. The band
    at $\pm20\%$ is just to guide the eye.
    \protect\label{f:distributions}}
\end{figure}

Background-subtracted variable distributions in data are compared to
the corresponding distributions in the MC simulation. In
Fig.~\ref{f:distributions} the distributions for the \pt\ of the
subleading muon (the muon with the lower \pt) and the \fls\ of the
\Bp\ meson are shown as examples. The MC simulation provides a
reasonable description of the data; the remaining discrepancies are
fully accounted for in the systematic uncertainty.  The pileup
dependence of the HLT tracking in Run 2 affects the normalization
sample stronger than the signal sample because of the displacement
requirement in the \bupsikp\ HLT path. This is corrected for with an
offline reweighting depending on the number of reconstructed PV and
\flsxy. The remaining systematic error from this corrections is
estimated to be 6\% for 2016A and 5\% for 2016B for the branching
fraction measurement. For the \bsmm\ effective lifetime measurement, a
systematic error of $0.07\ps$ is estimated. This is the second-largest
systematic uncertainty for both results.

\begin{figure}[!htb]
  \centerline{
    \includegraphics[width=0.52\textwidth,trim=0 3 0 0]{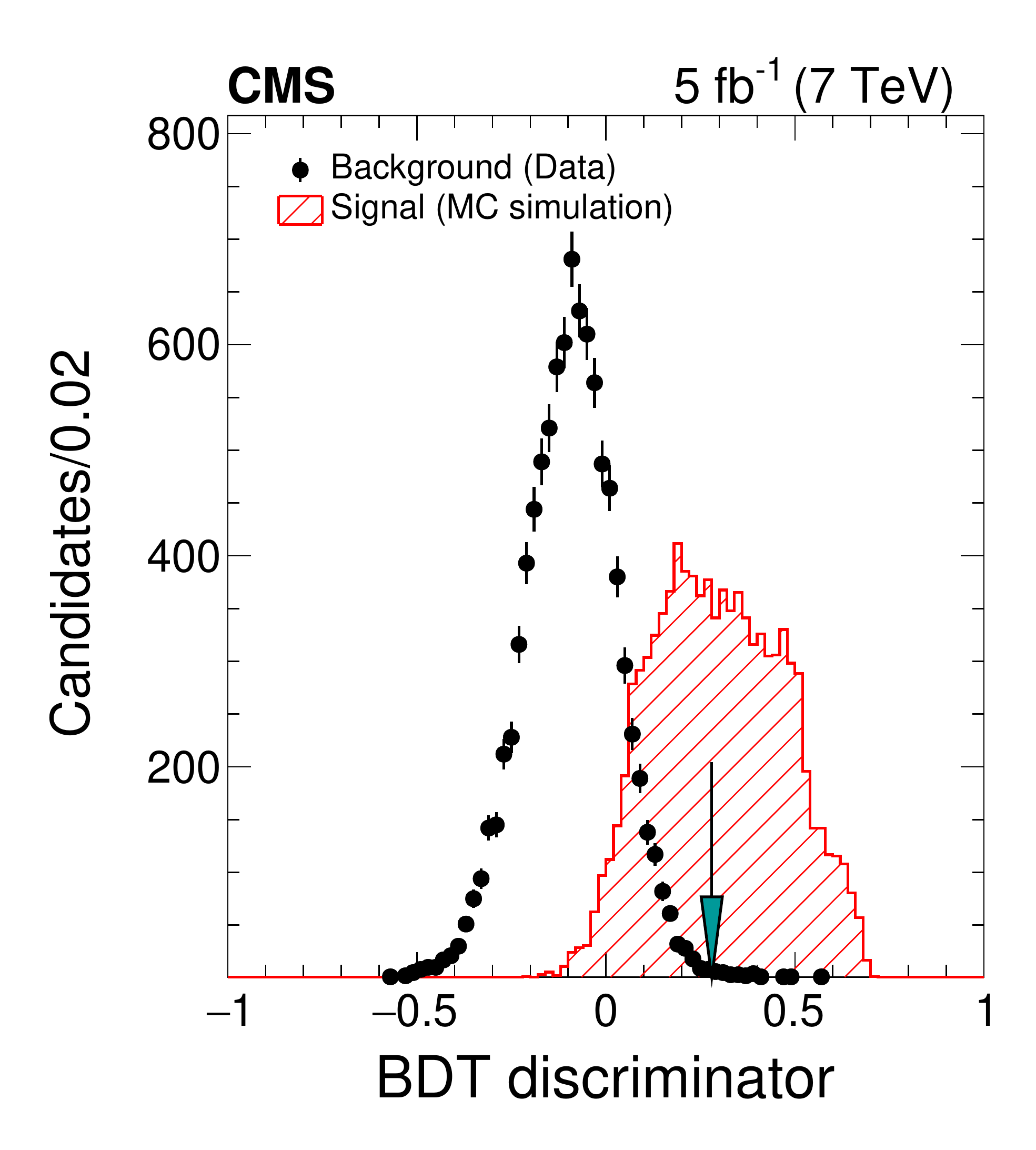}
    \hfill
    \includegraphics[width=0.5\textwidth]{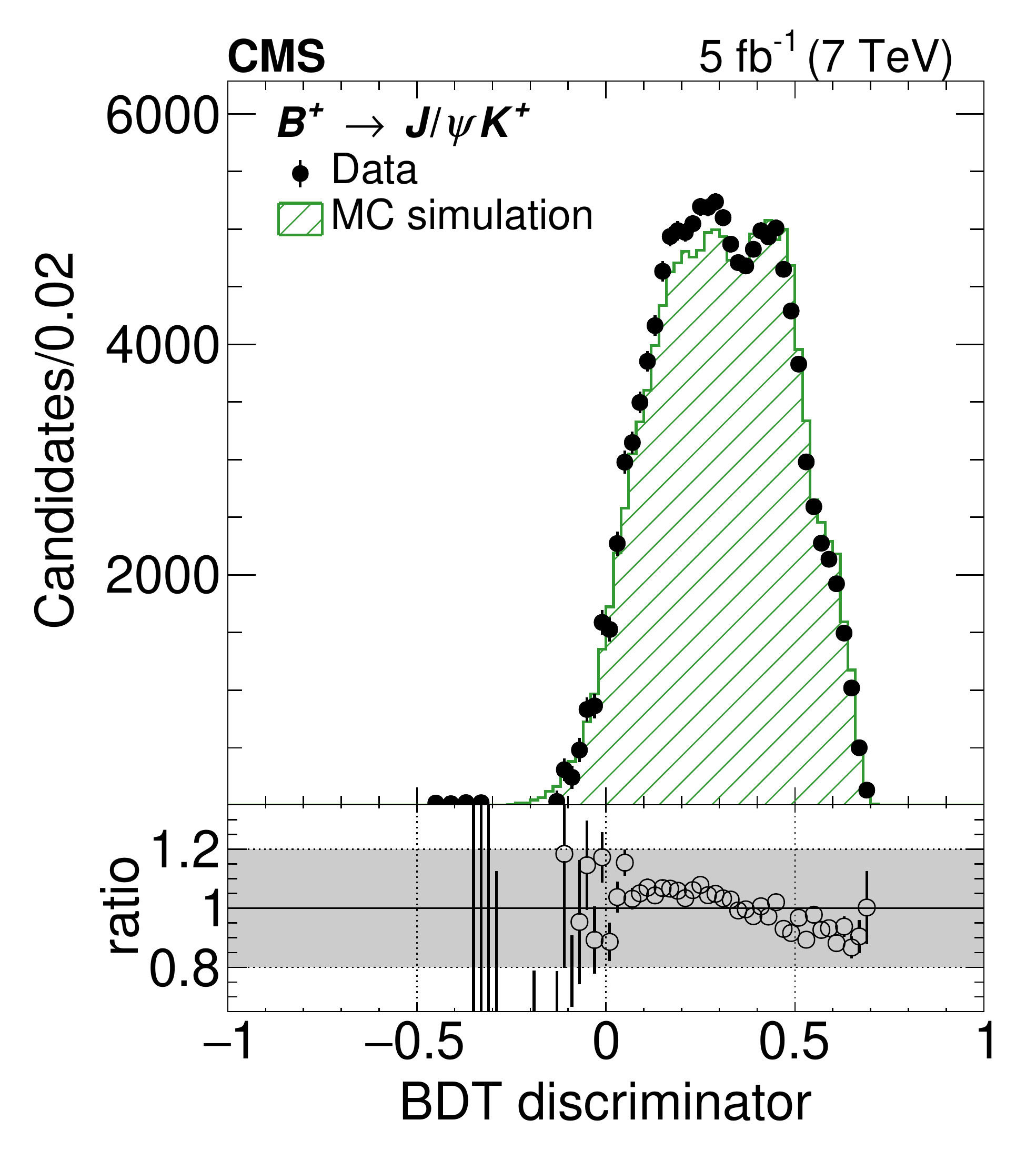}
  }
  \vspace*{8pt}
  \caption{Comparison of the BDT discriminator response for (left)
    dimuons in data (background from the invariant mass sideband $5.45
    < \mll < 5.9\gev$) and MC simulation (\bsmm\ signal) and (right)
    \bupsikp\ candidates in data and MC simulation.  The left plot
    illustrates the rejection power of the BDT against (combinatorial)
    background, the right plot shows the level of agreement between
    data and MC simulation, an ingredient in the determination of the
    systematic error of the analysis efficiency. The arrow in the left
    plot indicates the category boundary of the BDT discriminator
    range used in the $\cbf(\bsmm)$ result determination
    (cf.~Table~\ref{t:bdtcuts} below).  \protect\label{f:bdtresponse}}
\end{figure}

In Fig.~\ref{f:bdtresponse} the BDT discriminator response is shown
for dimuon candidates, illustrating the background rejection, and
for \bupsikp\ candidates, illustrating the agreement between data and MC
simulation. The systematic error of the selection efficiency is
estimated from the double ratio

\begin{equation}
D =\frac{\left[\frac{\varepsilon(\bupsik)}{\varepsilon(\bspsiphi)}\right]_{\text{data}}
}{\left[\frac{\varepsilon(\bupsik)}{\varepsilon(\bspsiphi)}\right]_{\text{MC}}},
\end{equation}
where the control sample \bspsiphi\ is used as a placeholder for the
signal sample. Depending on the channel, this systematic error varies
between 5\% and 10\%. It constitutes the largest contribution to the
overall systematic error for the branching fraction. The selection
efficiency depends on the unknown true effective lifetime
because of the displacement and isolation criteria. This uncertainty
of 1--3\%, depending on data-taking period and analysis channel, is
estimated with $\Delta \equiv [\varepsilon_{\text{tot}}(\tau_{\BsH}) -
  \varepsilon_{\text{tot}}(\tau_{\BsL})]/\sqrt{12}$ using simulated
samples with different effective lifetimes.

The mixture of \bbbar\ quark production processes in the MC simulation
is not necessarily the same as in data. At leading order three
processes contribute to heavy quark production in $pp$ collisions:
gluon splitting, gluon-gluon fusion, and flavor excitation. The $b$
quarks from gluon splitting are closer together in phase space than
for the other two processes, where the two $b$ quarks tend to be
back-to-back in the transverse plane. Therefore a \bsmm\ decay from a
\Bs\ meson produced in gluon splitting will be, on average, less
isolated than from a \Bs\ meson from gluon-gluon fusion. Therefore a
mismatch of the production process mixture in data and simulation will
imply a systematic uncertainty on the selection efficiency. The
mixture of production processes is studied by combining a
\bupsikp\ (\bspsiphi) candidate with another muon $\mu_3$ (assumed to
originate from the semileptonic decay of the other $b$ hadron in the
event) and studying the $\Delta R(B,\mu_3)$ distribution. Fitting
templates from MC simulation for gluon-splitting and the sum of
gluon-gluon fusion plus flavor excitation to the data allows an
estimate of the systematic uncertainty for the efficiency ratio at the
3\% level.

As a cross check, the effective\footnote{This is not an absolute
  measurement of $\cbf(\bspsiphi)$ because of \fsfu\ in
  Eq.~\ref{eq:schema}. However, the point of this study is not the
  absolute measurement but rather the study of the stability of this
  quantity in different channels, data-taking periods, and detector
  configurations.} branching fraction $\cbf(\bspsiphi)$ was determined
with an approach equivalent to Eq.~\ref{eq:schema}, where the number
of \bspsiphi\ decays is used for $N_{\Bs}$, for all channels in all
data-taking periods. The standard deviation of these eight
measurements is 4\%, smaller than the combination of the systematic
uncertainties due to analysis efficiency, tracking efficiency (the
decay \bspsiphi\ has one additional kaon track compared to \bupsikp),
hadronization uncertainties (\Bs\ vs.~\Bu), and yield
determinations. Therefore we conclude that the systematic error is not
underestimated.

Efficiency-corrected yield ratios for \bupsikp, \bspsiphi, and
\bdpsikstar\ (with $\Kstarz\to\Kp\pim$) decays are studied in the range
$10< \ptb < 100\gev$ and $0 < |\etab| < 2.2$ to estimate a possible
dependence on these kinematic variables. No significant slope is
observed vs.~\ptb\ or $\etab$ for \fsfu, \fsfd, or \fdfu\ (as a control
measurement).

\section{Results}
For the determination of the results, the per-channel event samples
are further subdivided into mutually exclusive categories of different
signal-to-background ratios by introducing high- and low-BDT
categories in the BDT discriminator distributions. This is illustrated
in Fig.~\ref{f:bdtresponse} (left), where the arrow shows the
boundary between these categories. This categorization was optimized
separately for the branching fraction measurement and the effective
lifetime measurements. Table~\ref{t:bdtcuts} provides the category
boundaries. They vary over the channels because the BDT configuration
has been optimized independently for each channel and results in
different BDT discriminator distributions.

\begin{table}[!htb]
  \tbl{BDT discriminator category boundaries for the branching
    fraction measurement (left part) and the effective lifetime
    measurements (right part). These boundaries are illustrated in
    Fig.~\ref{f:bdtresponse} (left). In the 2011 data-taking period,
    there is no low-BDT category because of the limited number of
    events.}{
    \begin{tabular}{@{}crrrr@{}}
      \toprule
      &\multicolumn{2}{c}{branching fraction measurement}  &\multicolumn{2}{c}{effective lifetime measurement}\\
      &central &forward &central &forward\\
      \toprule
      2011  &\{0.28, 1\}       &\{0.21, 1\} &\{0.22, 1\} &\{0.19, 1\}\\
      2012  &\{0.27, 0.35, 1\} &\{0.23, 0.32, 1\} &\{0.32, 1\} &\{0.32, 1\}\\
      2016A &\{0.19, 0.30, 1\} &\{0.19, 0.30, 1\} &\{0.22, 1\} &\{0.30, 1\}\\
      2016B &\{0.18, 0.31, 1\} &\{0.23, 0.38, 1\} &\{0.22, 1\} &\{0.29, 1\}\\
      \botrule
    \end{tabular}
    \label{t:bdtcuts}
  }
\end{table}

The branching fractions $\cbfb(\bsmm)$ and $\cbf(\bdmm)$ are
determined with a 3D unbinned extended maximum likelihood fit to the
dimuon invariant mass distribution, the relative mass resolution, and
the binary distribution for the dimuon pairing configuration $\calc$
($\calc = \pm1$ for the two muons bending towards or away from each
other, respectively). While the first two variables have been used
already in the past~\cite{Chatrchyan:2013bka}, the last variable was
added as a protection against a possible underestimation of the
\bhh\ background (critical for the \bdmm\ measurement). Such an
underestimate is possible because the \bhh\ contribution is determined
under the assumption that the dimuon fake rate is the product of the
single muon fake rates. While there is no evidence for such an
underestimation for $\calc = -1$ with the muons bending away from each
other, the situation is less clear for $\calc = +1$ where the muon
tracks can be close together in the muon system. Since the effect is
different for the two configurations, the $\calc$ distribution is
introduced. Its shape is taken from MC simulation for the signal and
all background components. In the fit, a scale factor is used to
correct the expected background component yields for the case $\calc =
+1$. A second, independent approach to control such an underestimate
is to strongly reduce the muon misidentification probability (in this
analysis, the misidentification probability was reduced by about 50\%
compared to the previous analysis~\cite{Chatrchyan:2013bka}).

The signal probability density functions (PDFs) are based on a Crystal
Ball function~\cite{bib-crystalball} for the invariant mass and a
nonparametric kernel estimator~\cite{Cranmer:2000du} based on Gaussian
kernels for the relative mass resolution. Table~\ref{t:pdfs} provides
a summary of all PDFs, for the branching fraction fit and the
effective lifetime fit. The width of the Crystal Ball function is a
conditional parameter with linear dependence on the dimuon mass
resolution. All parameters, except for the signal yields, are fixed to
values obtained from the MC simulation. Differences in the mass scale,
studied with $\jpsi\to\mup\mun$ and $\OneS\to\mup\mun$ and
interpolated to $m_{\Bs}$, are taken into account by shifting the MC
mass distributions. The difference in the mass resolution between data
and MC simulation has an effect of less than $0.2\%$ on the final
results and is neglected.

\begin{table}[!htb]
  \tbl{Summary of the PDFs used for signal and background components
    in the unbinned maximum likelihood fit for (top part of the table)
    $\cbf(\bsmm)$ and $\cbf(\bdmm)$ as well as for (bottom part of the
    table) \tmm. Parameters that are floated in the fit are
    explicitly indicated (normalizations $N$, \tmm, and background
    polynomial parameters $p_0$ and $p_1$ and lifetime $t_0$). The
    other parameters, fixed to the values obtained in MC simulation,
    are not shown explicitly. Functions whose parameters are
    determined (and fixed) in the sideband are indicated with /SB. The
    function abbreviations are as follows: Crystall-Ball (CB),
    Gaussian (G), Gaussian kernel estimator (KEYS), Bernstein
    polynomial of first degree (BE), binary distribution (BD),
    exponential including resolution and efficiency modeling (Exp),
    and exponential including resolution (Exp').  The arrow $\to$
    indicates that a component is absorbed into the entry to the
    right.}{
    \begin{tabular}{@{}lrrrrrr@{}}
      \toprule
      Variable &$\Bs\to\mu\mu$ &$\Bz\to\mu\mu$ &$B\to hh$ &$B\to h \mu\nu$ &$B\to h \mu\mu$ &Combinatorial\\
      &&&&&&Background\\
      \toprule
      \mll     &CB($N$) &CB($N$) &CB+G &KEYS &KEYS &BE($p_0, p_1$)\\
      $\sigma(\mll)/\mll$ &KEYS &KEYS &KEYS &KEYS &KEYS &KEYS/SB\\
      \calc    &BD &BD &BD &BD &BD &BD/SB\\
      \botrule
      \mll     &CB($N$) &$\to$ &CB+G &$\to$ &G &BE($p_0, p_1$)\\
      $t$      &Exp($N$, \tmm)   &$\to$ &Exp &$\to$ &Exp &Exp'($N$, $t_o$)\\
      \botrule
    \end{tabular}
    \label{t:pdfs}
  }
\end{table}

The combinatorial background is modeled with a nonnegative Bernstein
polynomial of the first degree (basis polynomials $p_{0,1}(x) = 1-x$
  and $p_{1,1}(x) = x$ for $x\in[0,1]$) with floating parameters in the fit. Using an
exponential function instead changes the result by $2.3\%$ ($0.6\%$)
for \bsmm\ (\bdmm), which is included in the systematic
uncertainties. The relative mass resolution is modeled with a kernel
function, determined from the data invariant mass sideband.

The rare background components are grouped together according to the
number of muons in the final state (zero, one, or two muons). Each
group is the weighted sum of various components. The peaking
background \bhh\ combines all modes with $h\in\{\pi, K, p\}$, while
the semileptonic group $B\to h \mu \nu$ consists of \bdpimunu,
\bskmunu, and \lbpmunu. Finally, the group $B\to h \mup\mun$ includes
\bupimumu\ and \bdmumupz. The weights in the sum are the product of
the misidentification probabilities of each hadron (depending on
charge, \pt, and $\eta$), the decay mode branching fraction, the
analysis efficiency, and the trigger efficiency. Using
Eq.~\ref{eq:schema} with the known branching fractions and the
measured \bupsikp\ normalization yield, it is possible to predict
absolutely the expected yield per decay mode. To account for the
missing components in the two groups with one or two muons, a common
scaling factor is applied such that their sum plus the combinatorial
background, extrapolated from the sideband $5.45 < \mll < 5.9\gev$,
matches the event yield in data in the mass region $4.9 < \mll <
5.2\gev$. The trigger efficiency for these modes cannot be determined
easily because of possible correlations between the two (fake) muons
and the very limited sample size of the MC simulation where 1--2
hadrons are misidentified as fake muons. Studies based on samples with
alternative muon identification algorithms indicate that the trigger
efficiency for \bhh\ is $\approx$50\% of the signal trigger efficiency
while for the other groups (with at least one muon) it is at the same
level as for the signal. A relative systematic error of 100\% is
assigned to the \bhh\ trigger efficiency and this dominates the
overall systematic uncertainty of the \bhh\ yield. For $B\to h \mu
\nu$ and $B\to h \mup \mun$ the systematic error on the yield is about
15\%. In the fit, the rare background yields are constrained to the
expectations within these uncertainties.

\begin{figure}[!htb]
  \centerline{
    \includegraphics[width=0.45\textwidth]{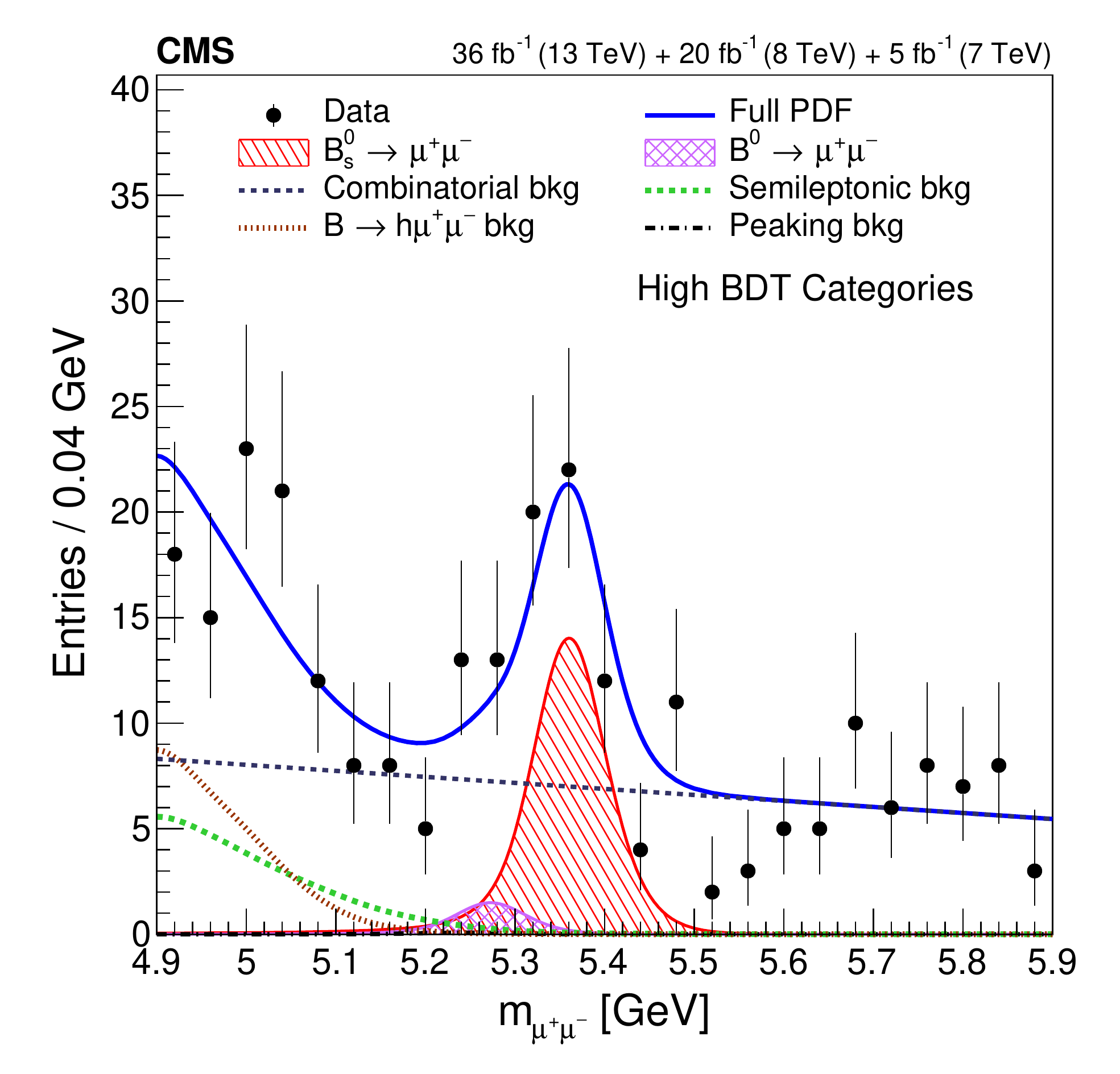}
    \hfill
    \includegraphics[width=0.45\textwidth]{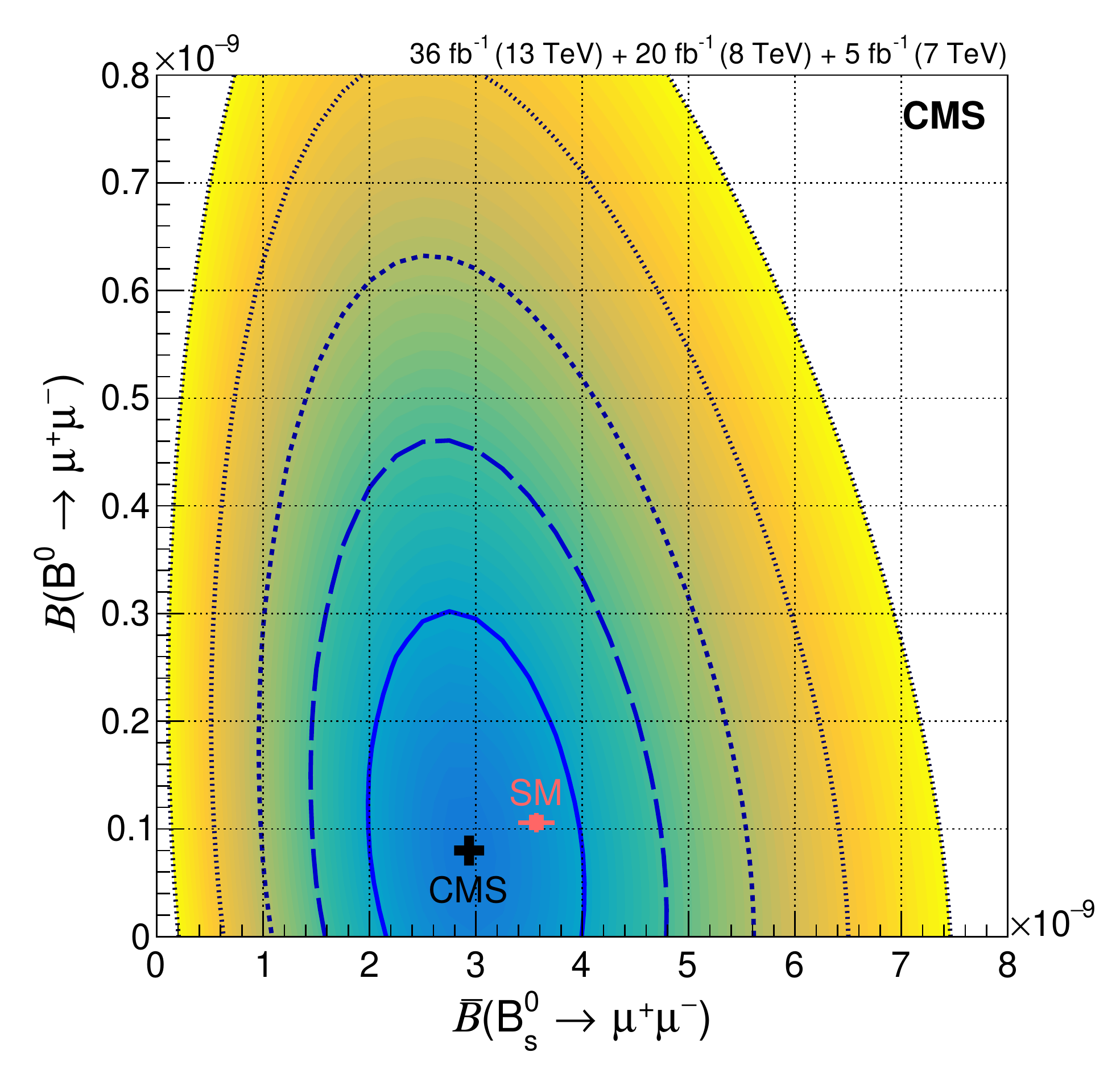}
  }
  \vspace*{8pt}
  \caption{(left) Invariant dimuon mass distribution for the high BDT
    categories (cf.~Table~\ref{t:bdtcuts}), with the fit projections
    overlaid. The peaking background is virtually invisible. (right)
    Likelihood contours of the branching fraction fit for
    $\cbfb(\bsmm)$ and $\cbf(\bdmm)$, with the best-fit value (black
    cross) and the SM expectation (red solid
    square).\protect\label{f:bfresults}}
\end{figure}

The only parameters of interest in the fit are $\cbfb(\bsmm)$ and
$\cbf(\bdmm)$. All other parameters are nuisance parameters and are
subject to Gaussian constraints, except for the rare background yields
where log-normal priors are used as constraints. In
Fig.~\ref{f:bfresults} (left) the mass distribution of the high BDT
categories is shown. The \bsmm\ signal is clearly visible. As a
consequence of the substantially improved muon identification
algorithm, the peaking background is virtually invisible---a
significant improvement compared to
Ref.~\refcite{Chatrchyan:2013bka}. The result of the fit to the data
in the 14 categories, as defined in Table~\ref{t:bdtcuts} (left part),
is
\begin{equation}
\cbfb(\bsmm) = \resObsBFBsmm,
\end{equation}
where the large statistical ($\pm0.6\times10^{-9}$) and small
systematical ($\pm0.3\times10^{-9}$) errors are combined into one
experimental uncertainty, and the second error is due to the
\fsfu\ uncertainty. Using Wilks' theorem~\cite{Wilks:1938dza}, the
observed (expected) significance amounts to $\sigObsBFBsmm\sigma$
($\sigExpBFBsmm\sigma$). The fit likelihood contours are shown in
Fig.~\ref{f:bfresults} (right); the correlation between the two
branching fractions is $-0.181$.  Summing over all BDT categories, a
total \bsmm\ signal yield of $61{}^{+15}_{-13}$ events is observed,
with an average \ptb\ of $17.2\gev$.  For $\cbfb(\bsmm)$ the
systematic error is dominated by the efficiency difference between
data and MC simulation and by the pileup dependent effects of the HLT
tracking in Run 2. All other components of the systematic error are
much less important.

The fit also determines $\cbf(\bdmm) = \resObsBFBdmm$ with an observed
(expected) significance of $\sigObsBFBdmm\,\sigma$
($\sigExpBFBdmm\,\sigma$). Because no significant result was expected
in this case, the primary result here is $\cbf(\bdmm) < \ulaBFBdmm
(\ulbBFBdmm)$ at $\ulacl (\ulbcl)\%$ confidence level, using the
CL$_{\mathrm s}$ method~\cite{Read2002,Junk1999} with the standard
LHC-type profiled likelihood. The corresponding expected upper limit
is $\cbf(\bdmm) < \ulaExpBdmm (\ulbExpBdmm)$, assuming no signal. The
systematic error for $\cbf(\bdmm)$ is similar as for $\cbfb(\bsmm)$,
with the exception of the rare background yields. However, given the
precision of the upper limits quoted, this difference has no numerical
impact.

For the determination of the \bsmm\ effective lifetime \tmm, two
independent fit frameworks were established, a 2D unbinned maximum
likelihood fit to the invariant mass and decay time distributions and
a 1D binned maximum likelihood fit to the decay time distribution
where the background is subtracted with the {\it sPlot\/}
method~\cite{Pivk:2004ty}. Prior to unblinding the data, the former
was chosen as the primary method based on its better median expected
performance. The fits are performed for the decay time range $1 < t <
11\ps$ in eight BDT categories provided in Table~\ref{t:bdtcuts}. The
decay time restriction is motivated by the very low selection
efficiency at smaller decay times $t$, because of the \fls\ and
isolation requirements, while the efficiency at large $t$ is strongly
decreasing in Run 2, due to HLT requirements on the muon impact
parameter. Concerns regarding the fit stability, given the small
expected number of signal events, motivated using eight instead of 14 BDT
categories.

\begin{figure}[!htb]
  \centerline{
    \includegraphics[width=0.45\textwidth]{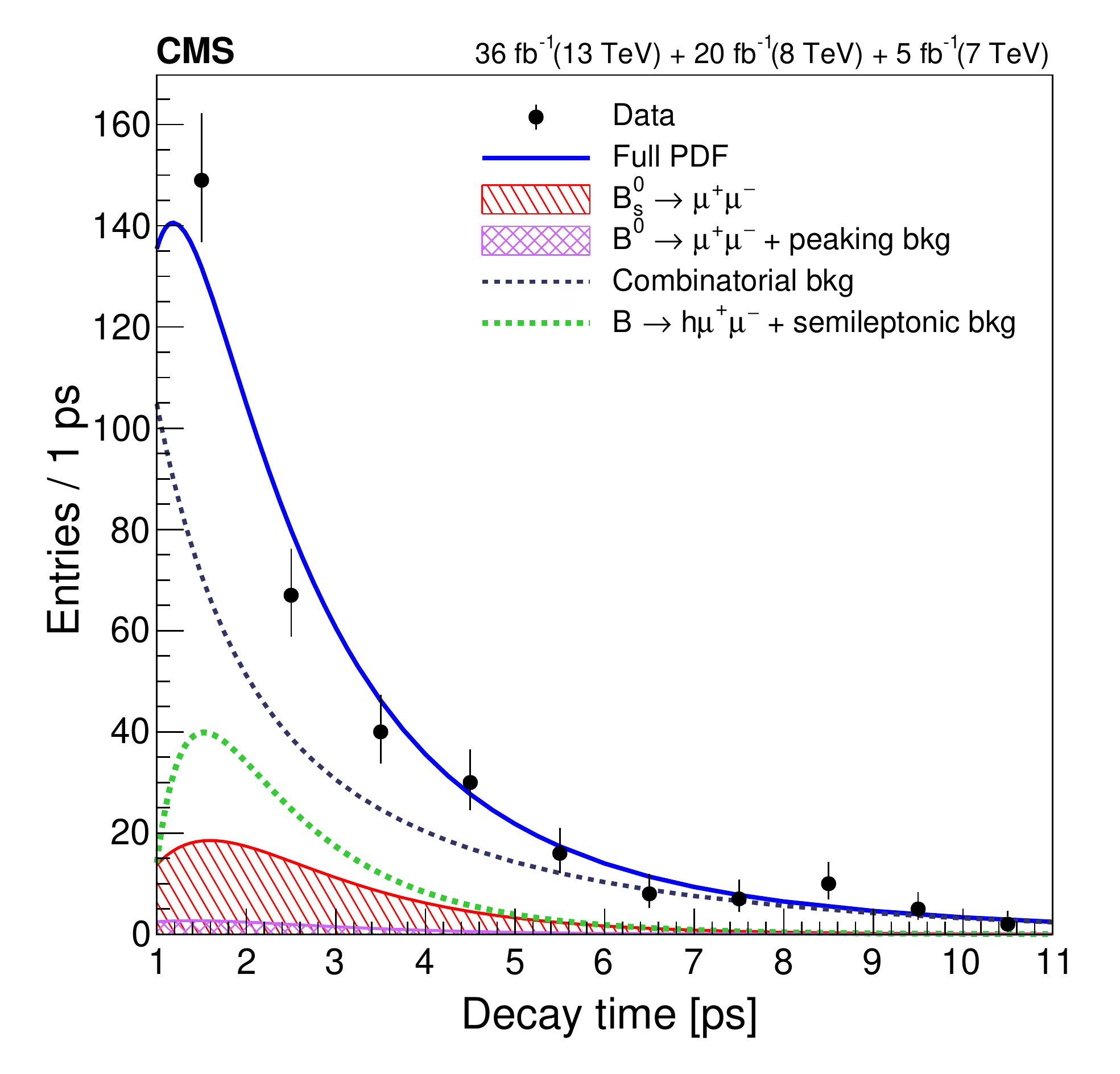}
    \hfill
    \includegraphics[width=0.45\textwidth]{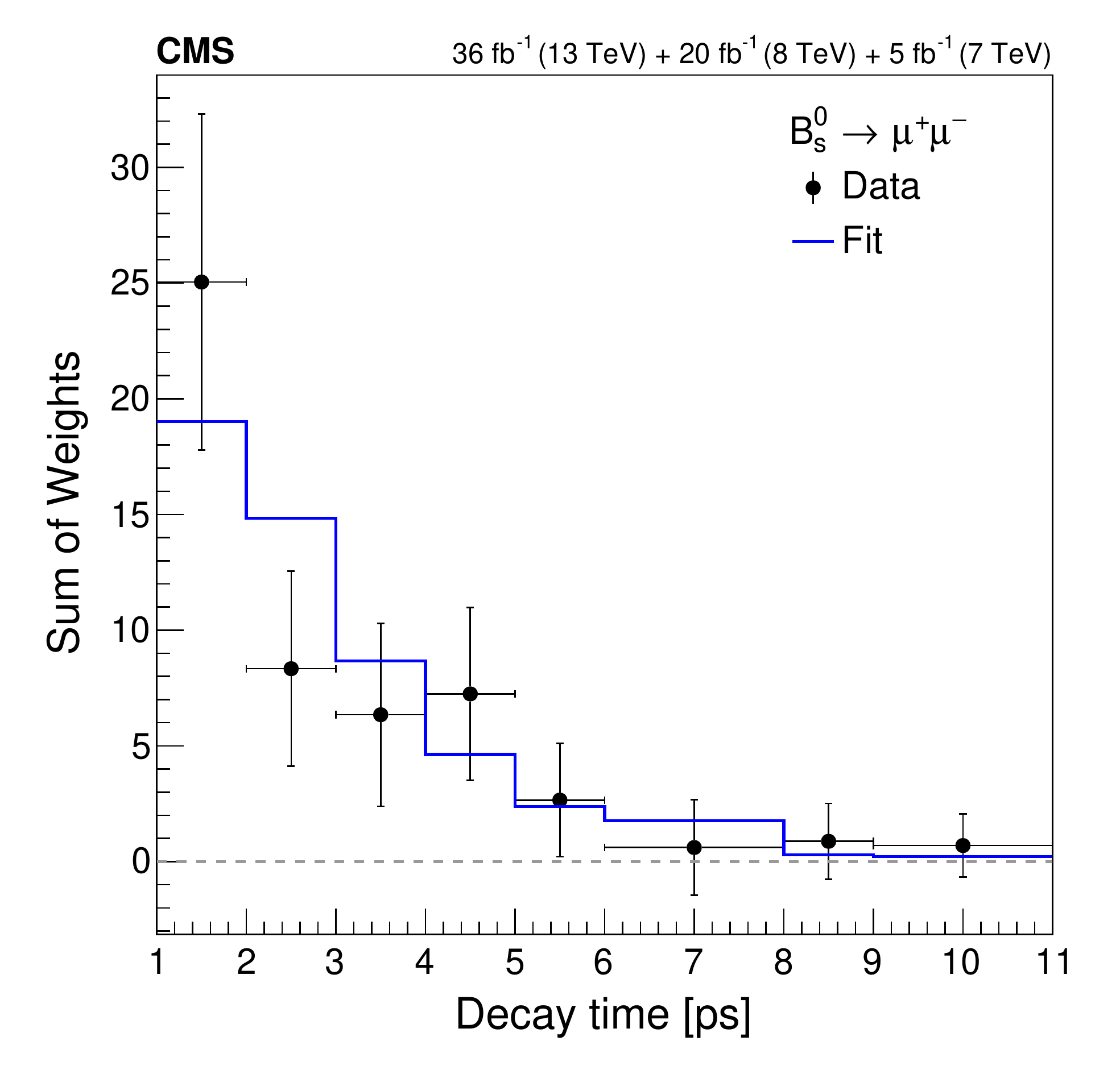}
  }
  \vspace*{8pt}
  \caption{Decay time distributions for (left) the 2D unbinned maximum
    likelihood fit and (right) the {\it sPlot\/} fit approach. The
    data combine all events in the eight BDT categories of
    Table~\ref{t:bdtcuts} (right part). In the left plot, the total
    fit is shown by the solid line, the signal by the single-hatched
    distribution, and the different background components by the
    broken lines and the cross-hatched distributions. In the right
    plot, only the signal component is shown as the background has
    been subtracted beforehand with the {\it sPlot\/} method (see text
    for details). \protect\label{f:tauresults}}
\end{figure}

The PDFs for the invariant mass of the 2D unbinned maximum likelihood
fit are very similar to those in the branching fraction fits with the
notable exception that \bdmm\ is included in the peaking background.
In the decay time PDFs, the exponential functions are convolved with
Gaussian functions to account for detector resolution. The decay time
efficiency is included in all fit components except for the
combinatorial background, because its PDF is modeled from data
directly. In the fit, the \bsmm\ effective lifetime \tmm, the signal
yield, and the parameters of the combinatorial background are floated
(cf.~Table~\ref{t:pdfs} for a summary). All other parameters are
constrained or fixed to the MC simulation values. There is no common
$\cbfb(\bsmm)$ constraint for the \bsmm\ signal yields in the eight
BDT categories. The result of the fit is
\begin{equation}
\tmm(\bsmm) = \resObsTauBsmm\,\ps,
\end{equation}
where the error combines the large statistical
(${}^{+0.60}_{-0.43}\ps$) and small systematic ($\pm0.09\ps$)
uncertainty. In Fig.~\ref{f:tauresults} (left) the decay time
distribution is shown, together with the fit results overlaid. The
observed errors are about one root-mean-square deviation larger than
expected (${}^{+0.39}_{-0.30}\ps$). This is attributed to fluctuations
in the small sample size.

The second determination of \tmm\ uses the complete model of the
branching fraction fit to determine the {\it sPlot\/} weights. An
exponential function, modified to include the channel-dependent
resolution and efficiency effects, is fit to the {\it sPlot\/}
distribution. Special care is applied to determine asymmetric
uncertainties and to reduce the bias due to the large bin widths. The
fit yields $\tmm = \resObsTauSplot\ps$, where a fit bias of $+0.09\ps$
has been corrected for. This bias, together with the dependence on the
Run 2 data-taking periods, is the largest systematic error in this
approach. The two determinations of the effective lifetime are
consistent with each other.

\section{Conclusions}
The CMS experiment has analyzed rare leptonic \bmm\ decays with
61\invfb\ of data collected in LHC Run 1 and 2016. The decay \bsmm\ is
observed with a significance of $\sigObsBFBsmm\,\sigma$ ($6.5\,\sigma$
expected). Its branching fraction is measured to be $\cbfb(\bsmm) =
\resObsBFBsmm$ and the \bsmm\ effective lifetime is determined to be
$\tmm = \resObsTauBsmm\ps$. Both results are limited by the small
signal sample. No evidence is found for \bdmm\ and an upper limit of
$\cbf(\bdmm) < \ulaBFBdmm$ (at 95\%\,confidence level) is
determined. These results are consistent with the SM. The branching
fraction results are also consistent with the previous CMS
analysis~\cite{Chatrchyan:2013bka} when restricting this analysis to
the Run 1 dataset.

\section*{Acknowledgments}
It is a pleasure to acknowledge the common effort and discussions on
\bmm\ decays over the past 15 years with many colleagues of the CMS
collaboration. Countless stimulating and instructive discussions with
Andreas Crivellin are very much appreciated.


\bibliographystyle{ws-mpla}
\bibliography{ws-mpla}

\begin{thebibliography}{10}

\bibitem{Bobeth:2013uxa}
C.~Bobeth, M.~Gorbahn, T.~Hermann, M.~Misiak, E.~Stamou and M.~Steinhauser,
  {\em Phys. Rev. Lett.} {\bf 112},   101801  (2014),
  \href{http://arxiv.org/abs/1311.0903}{{\ttfamily arXiv:1311.0903 [hep-ph]}}.

\bibitem{Hermann:2013kca}
T.~Hermann, M.~Misiak and M.~Steinhauser, {\em JHEP} {\bf 12},   097  (2013),
  \href{http://arxiv.org/abs/1311.1347}{{\ttfamily arXiv:1311.1347 [hep-ph]}}.

\bibitem{Bobeth:2013tba}
C.~Bobeth, M.~Gorbahn and E.~Stamou, {\em Phys. Rev. D} {\bf 89},   034023
  (2014), \href{http://arxiv.org/abs/1311.1348}{{\ttfamily arXiv:1311.1348
  [hep-ph]}}.

\bibitem{Beneke:2017vpq}
M.~Beneke, C.~Bobeth and R.~Szafron, {\em Phys. Rev. Lett.} {\bf 120},   011801
   (2018), \href{http://arxiv.org/abs/1708.09152}{{\ttfamily arXiv:1708.09152
  [hep-ph]}}.

\bibitem{Beneke:2019slt}
M.~Beneke, C.~Bobeth and R.~Szafron, {\em JHEP} {\bf 10},   232  (2019),
  \href{http://arxiv.org/abs/1908.07011}{{\ttfamily arXiv:1908.07011
  [hep-ph]}}.

\bibitem{Cabibbo:1963yz}
N.~Cabibbo, {\em Phys. Rev. Lett.} {\bf 10}, 531  (1963).

\bibitem{Kobayashi:1973fv}
M.~Kobayashi and T.~Maskawa, {\em Prog. Theor. Phys.} {\bf 49}, 652  (1973).

\bibitem{pdg2018}
{Particle Data Group}, M.~Tanabashi {\em et~al.}, {\em Phys. Rev. D} {\bf 98},
   030001  (2018).

\bibitem{DeBruyn:2012wj}
K.~De~Bruyn, R.~Fleischer, R.~Knegjens, P.~Koppenburg, M.~Merk and N.~Tuning,
  {\em Phys. Rev. D} {\bf 86},   014027  (2012),
  \href{http://arxiv.org/abs/1204.1735}{{\ttfamily arXiv:1204.1735 [hep-ph]}}.

\bibitem{Aoki:2019cca}
Flavour Lattice Averaging Group Collaboration, S.~Aoki {\em et~al.}, {\em Eur.
  Phys. J. C} {\bf 80},   113  (2020),
  \href{http://arxiv.org/abs/1902.08191}{{\ttfamily arXiv:1902.08191
  [hep-lat]}}.

\bibitem{Aaboud:2018mst}
ATLAS Collaboration, M.~Aaboud {\em et~al.}, {\em JHEP} {\bf 04},   098
  (2019), \href{http://arxiv.org/abs/1812.03017}{{\ttfamily arXiv:1812.03017
  [hep-ex]}}.

\bibitem{Chatrchyan:2013bka}
CMS Collaboration, S.~Chatrchyan {\em et~al.}, {\em Phys. Rev. Lett.} {\bf
  111},   101804  (2013), \href{http://arxiv.org/abs/1307.5025}{{\ttfamily
  arXiv:1307.5025 [hep-ex]}}.

\bibitem{Aaij:2013aka}
LHCb Collaboration, R.~Aaij {\em et~al.}, {\em Phys. Rev. Lett.} {\bf 111},
  101805  (2013), \href{http://arxiv.org/abs/1307.5024}{{\ttfamily
  arXiv:1307.5024 [hep-ex]}}.

\bibitem{Aaij:2017vad}
LHCb Collaboration, R.~Aaij {\em et~al.}, {\em Phys. Rev. Lett.} {\bf 118},
  191801  (2017), \href{http://arxiv.org/abs/1703.05747}{{\ttfamily
  arXiv:1703.05747 [hep-ex]}}.

\bibitem{Sirunyan:2019xdu}
CMS Collaboration, A.~M. Sirunyan {\em et~al.}, {\em JHEP} {\bf 04},   188
  (2020), \href{http://arxiv.org/abs/1910.12127}{{\ttfamily arXiv:1910.12127
  [hep-ex]}}.

\bibitem{Abazov:2004dj}
D0 Collaboration, V.~Abazov {\em et~al.}, {\em Phys.\ Rev.\ Lett.} {\bf 94},
  071802  (2005), \href{http://arxiv.org/abs/hep-ex/0410039}{{\ttfamily
  arXiv:hep-ex/0410039}}.

\bibitem{Aaij:2019eej}
LHCb Collaboration, R.~Aaij {\em et~al.}, {\em Phys. Rev. Lett.} {\bf 124},
  122002  (2020), \href{http://arxiv.org/abs/1910.09934}{{\ttfamily
  arXiv:1910.09934 [hep-ex]}}.

\bibitem{Aad:2015cda}
ATLAS Collaboration, G.~Aad {\em et~al.}, {\em Phys. Rev. Lett.} {\bf 115},
  262001  (2015), \href{http://arxiv.org/abs/1507.08925}{{\ttfamily
  arXiv:1507.08925 [hep-ex]}}.

\bibitem{Aaij:2019pqz}
LHCb Collaboration, R.~Aaij {\em et~al.}, {\em Phys. Rev. D} {\bf 100},
  031102  (2019), \href{http://arxiv.org/abs/1902.06794}{{\ttfamily
  arXiv:1902.06794 [hep-ex]}}.

\bibitem{Chatrchyan:2008zzk}
CMS Collaboration, S.~Chatrchyan {\em et~al.}, {\em JINST} {\bf 3},   S08004
  (2008).

\bibitem{Khachatryan:2010pw}
CMS Collaboration, V.~Khachatryan {\em et~al.}, {\em Eur. Phys. J. C} {\bf 70},
    1165  (2010), \href{http://arxiv.org/abs/1007.1988}{{\ttfamily
  arXiv:1007.1988 [physics.ins-det]}}.

\bibitem{CMS-DP-2018-050}
CMS Collaboration, {CMS Collaboration}, {\em Tracking {POG} results for pion
  efficiency with the {$\Dstarp$} meson using data from 2016 and 2017}, CMS
  Detector Performance Note CMS-DP-2018-050  (2018).

\bibitem{Chatrchyan:2012xi}
CMS Collaboration, S.~Chatrchyan {\em et~al.}, {\em JINST} {\bf 7},   P10002
  (2012), \href{http://arxiv.org/abs/1206.4071}{{\ttfamily arXiv:1206.4071
  [physics.ins-det]}}.

\bibitem{Sirunyan:2018}
CMS Collaboration, A.~M. Sirunyan {\em et~al.}, {\em JINST} {\bf 13},   P06015
  (2018), \href{http://arxiv.org/abs/1804.04528}{{\ttfamily arXiv:1804.04528
  [physics.ins-det]}}.

\bibitem{Khachatryan:2016bia}
CMS Collaboration, V.~Khachatryan {\em et~al.}, {\em JINST} {\bf 12},   P01020
  (2017), \href{http://arxiv.org/abs/1609.02366}{{\ttfamily arXiv:1609.02366
  [physics.ins-det]}}.

\bibitem{Hocker:2007ht}
H.~Voss, A.~H{\"o}cker, J.~Stelzer and F.~Tegenfeldt, {\it {TMVA}, the toolkit
  for multivariate data analysis with {ROOT}}, in {\em XIth International
  Workshop on Advanced Computing and Analysis Techniques in Physics Research
  (ACAT)\/},  (2007).
\newblock p.~40.
\newblock \href{http://arxiv.org/abs/physics/0703039}{{\ttfamily
  arXiv:physics/0703039}}.
\newblock {[PoS(ACAT)040]}.

\bibitem{bib-crystalball}
M.~J. Oreglia, {\it A study of the reactions $\psi' \to \gamma\gamma \psi$},
  PhD thesis, Stanford University, (1980).
\newblock {SLAC-R-236, UMI-81-08973. See Appendix D}.

\bibitem{Cranmer:2000du}
K.~S. Cranmer, {\em Comput. Phys. Commun.} {\bf 136},   198  (2001),
  \href{http://arxiv.org/abs/hep-ex/0011057}{{\ttfamily arXiv:hep-ex/0011057
  [hep-ex]}}.

\bibitem{Wilks:1938dza}
S.~S. Wilks, {\em Annals Math. Statist.} {\bf 9},  ~60  (1938).

\bibitem{Read2002}
A.~L. Read, {\em J. Phys. G} {\bf 28},   2693  (2002).

\bibitem{Junk1999}
T.~Junk, {\em Nucl. Instrum. Meth. A} {\bf 434},   435  (1999),
  \href{http://arxiv.org/abs/hep-ex/9902006}{{\ttfamily arXiv:hep-ex/9902006}}.

\bibitem{Pivk:2004ty}
M.~Pivk and F.~R. Le~Diberder, {\em Nucl. Instrum. Meth. A} {\bf 555},   356
  (2005), \href{http://arxiv.org/abs/physics/0402083}{{\ttfamily
  arXiv:physics/0402083 [physics.data-an]}}.

\end{thebibliography}

\end{document}